\documentstyle[aps,epsf,manuscript,eqsecnum]{revtex}
\input epsf
\def\UU{Y_U Y_U^\dagger}
\def\DD{Y_D Y_D^\dagger}
\def\EE{Y_E Y_E^\dagger}
\def\dudt{16\pi^2{dY_U \over dt}}
\def\dddt{16\pi^2{dY_D \over dt}}
\def\dedt{16\pi^2{dY_E \over dt}}
\def\dvdt{16\pi^2{dv \over dt}}
\def\dvudt{16\pi^2{dv_u \over dt}}
\def\dvddt{16\pi^2{dv_d \over dt}}
\def\dMudt{16\pi^2{dM_U \over dt}}
\def\dMddt{16\pi^2{dM_D \over dt}}
\def\dMedt{16\pi^2{dM_E \over dt}}
\def\dudt{16\pi^2{dY_U \over dt}}
\def\dMus{d{\hat{M}_U}^2 \over dt}
\def\dMds{d{\hat{M}_D}^2 \over dt}
\def\dMes{d{\hat{M}_E}^2\over dt}
\def\Mus{{\hat{M}_U}^2}
\def\Mds{{\hat{M}_D}^2}
\def\Mes{{\hat{M}_E}^2}
\def\uudot{L_U^\dagger\dot{L}_U}
\def\dddot{L_D^\dagger\dot{L}_D}
\def\eedot{L_E^ \dagger\dot{L}_E}
\def\dmudt{16\pi^2{dm_u\over dt}}
\def\dmcdt{16\pi^2{dm_c \over dt}}
\def\dmtdt{16\pi^2{dm_t \over dt}}
\def\dmjdt{16\pi^2{dm_j \over dt}}
\def\dmbdt{16\pi^2{dm_b \over dt}}
\def\dmtaudt{16\pi^2{dm_{\tau} \over dt}}
\def\dgdt{16\pi^2{dg_i \over dt}}
\def\be{\begin{equation}}
\def\ee{\end{equation}}
\def\ba{\begin{eqnarray}}
\def\ea{\end{eqnarray}}
\def\br{\begin{array}}
\def\er{\end{array}}
\begin{document}
\draft
\title{New Formulas and Predictions for Running Fermion Masses at Higher
Scales in SM, 2HDM, and MSSM}
\author{C.R.~Das\footnote{crdas@email.com} and M.K.~Parida
\footnote{mparida@dte.vsnl.net.in}}
\address{Physics Department, North-Eastern Hill University\\
Shillong 793022, India}
\date{October 1, 2000}
\maketitle
\makeatletter
\begin{abstract}
Including contributions of scale-dependent vacuum expectation values, we
derive new analytic formulas and obtain substantially different numerical
predictions for the running  masses of quarks and charged-leptons at higher
scales in the SM, 2HDM and MSSM. These formulas exhibit significantly
different behaviours with respect to their dependence on gauge and Yukawa
couplings than those derived earlier. At one-loop level the masses of the
first two generations are found to be independent of Yukawa couplings of the
third generation in all the three effective theories in the small mixing
limit. Analytic formulas are also obtained for running $\tan\beta(\mu)$ in
2HDM and MSSM. Other numerical analyses include study of the third generation
masses at high scales  as functions of low-energy values of $\tan\beta$ and
SUSY scale $M_S=M_Z-10^4$ GeV.
\end{abstract}
\narrowtext
\par
\section{Introduction}
\label{sec1}
\par
One of the most attractive features of current investigations in gauge
theories is the remarkable unification of the gauge couplings of the standard
model (SM) at the SUSY GUT scale, $M_U=2\times 10^{16}$ GeV, when extrapolated
through the minimal supersymmetric standard model (MSSM) \cite{1}. Although
the nonsupersymmetric standard model (SM), or the two-Higgs doublet model
(2HDM) do not answer the question of gauge hierarchy, unification of the gauge
couplings is also possible at the corresponding GUT scales when they are
embedded in nonSUSY theories like $SO(10)$ and the symmetry breaking takes
place in two steps with left-right models as intermediate gauge symmetries
\cite{2}. Grand unification of gauge couplings of the SM in single-step
breakings of GUTs has also been observed when the grand desert contains
additional scalar degrees of freedom \cite{3} and the minimal example is a
$\xi(3,0,8)$ of SM contained in $\underline{75}\subset SU(5)$ or
$\underline{210}\subset SO(10)$ with mass $M_{\xi}=10^{11}-10^{13}$ GeV
\cite{4}. Unification of gauge couplings in nonSUSY $SO(10)$ has been
demonstrated with relatively large GUT-threshold effects \cite{5}.
Yukawa coupling unification at the intermediate scale has also been
observed in nonSUSY $SO(10)$ with 2HDM as the weak scale effective gauge
theory \cite{6}. Apart from unity of forces at high scales, SM, 2HDM and MSSM
have tremendous current importance as effective theories as they emerge from
a large class of fundamental theories.
\par
Recent experimental evidences in favour of neutrino masses and mixings have
triggered an outburst of models many of which require running masses and
mixings of quarks and charged-leptons at high scales as inputs for obtaining
predictions in the neutrino sector \cite{7}. The running masses are not only
essential at the weak scale, but they are also required at the intermediate
and the GUT scales in order to testify theories based upon quark-lepton
unification with different Yukawa textures and for providing unified
explanation of all fermion masses \cite{8,9,10,11,12}. Quite recently
extrapolation of running masses and couplings to high scales have been
emphasized as an essential requirement for testing more fundamental theories
\cite{12}.
\par
In a recent paper one of us (M.K.P) and Purkayastha \cite{13} have obtained
new analytic formulas and numerical estimations for the fermion masses at
higher scales in MSSM including contributions of scale-dependent vacuum
expectation values (VEVs) where the SUSY scale $(M_S)$ was assumed to be close
to the weak scale $(M_S\approx M_Z)$. In this paper we extend such
investigations to SM, 2HDM and MSSM with the SUSY scale $M_S\ge O$ (TeV).
\par
It is also possible that in a different renormalisation scheme, similar to
that formulated by Sirlin et.~al.~\cite{14}, the VEVs themselves do not run
when they are expressed in terms of physical parameters defined on the mass
shell. This makes it possible to avoid separate running of the VEVs and Yukawa
couplings, but have just the fermion masses directly as running quantities.
While it would be quite interesting to examine the consequences of such a
scheme, the purpose of the present and the recent works \cite{13} is to
address the outcome of the most frequently exploited renormalisation scheme
where the Yukawa couplings and the VEVs run separately
\cite{15,16,17,18,19,20,21,22,23}.
\par
This paper is organised in the following manner. In Sec.~\ref{sec2} we cite
examples where running VEVs have been exploited by a number of authors and
state relevant renormalisation group equations (RGEs). In Sec.~\ref{sec3} we
derive analytic formulas. In Sec.~\ref{sec4} we show how the formulas derived
earlier for MSSM are modified when $M_S \gg M_Z$. Numerical predictions at
higher scales are reported in Sec.~\ref{sec5}. Summary and conclusions are
stated in Sec.~\ref{sec6}.
\par
\section{RGEs for Couplings and Vacuum Expectation Values}
\label{sec2}
\par
After the pioneering discovery of $b-\tau$ unification at the nonSUSY $SU(5)$
GUT scale \cite{24}, a number of theoretical investigations have been made to
examine the behaviour of Yukawa couplings and running masses at higher scales.
Following the frequently exploited renormalisation scheme
\cite{15,16,17,18,19,20,21,22,23} where the Yukawa couplings and the VEVs run
separately, the running Dirac mass of the fermion `$a$' is defined as,
\par
\be M_a(\mu)=Y_a(\mu) v_a(\mu)\label{eq1}\ee
Then the running of $M_a(\mu)$ is governed by the RGEs of $Y_a(\mu)$ and
$v_a(\mu)$ both. To cite some examples, Grimus \cite{21} has derived
approximate analytic formulas in SM for all values of $\mu$ extending upto
the nonSUSY $SU(5)$ GUT scale utilising the corresponding scale-dependent VEV.
In the discovery of fixed point of Yukawa couplings, Pendleton and Ross
\cite{22} have exploited the RGE of the SM Higgs VEV to derive the RGEs of the
running masses from $\mu=M_W-M_{\rm GUT}$. Anomalous dimensions occurring in
the RGEs of respective VEVs have been explicitly derived and stated up to
two-loops by Arason et.~al.~\cite{15,16} and by Castano, Pirad and Ramond
\cite{17} for SM and MSSM. While investigating renormalisation of the neutrino
mass operator, Babu, Leung and Pantaleone \cite{23} have derived RGE for
$\tan\beta(\mu)$ in a class of 2HDM as a consequence of running VEVs in the
model. More recently Balzeleit et.~al.~\cite{19} have utilised the RGE of the
VEV in SM to determine running masses for $\mu = M_W-10^{10}$ GeV. Cvetic,
Hwang and Kim \cite{20} have derived RGEs for the VEVs in 2HDM and utilised
them to obtain running quark-lepton masses at high scales and also investigate
suppression of flavour changing neutral current in the model. Most recently
the RGEs of running VEVs have been utilised by one of us (MKP) and Purkayastha
\cite{13} who have obtained new analytic formulas and numerical estimations of
the fermion masses at higher scales taking the SUSY scale $M_S\approx M_Z$.
\par
We consider only the class of 2HDM where $\Phi_u$ gives masses to up-quarks
and $\Phi_d$ to down-quarks and charged-leptons. For the sake of simplicity
we ignore neutrino mass in the present paper which will be addressed
separately. Our definitions and conventions for the Yukawa couplings and
masses are governed by the following Yukawa Lagrangian (Superpotential) in SM
or 2HDM (MSSM) and the corresponding VEVs of Higgs scalars,
\\
\underline{SM}
$${\cal L}_Y=\overline{Q}_LY_U\tilde\Phi U_R+\overline{Q}_LY_D\Phi D_R+
\overline{l}_LY_E\Phi E_R+h.c.$$
\be\langle\Phi^0(\mu)\rangle=v(\mu)\label{eq2}\ee
\underline{2HDM, MSSM}
$${\cal L}_Y=\overline{Q}_LY_U\Phi_u U_R+\overline{Q}_LY_D\Phi_d D_R+
\overline{l}_LY_E\Phi_d E_R+h.c.$$
$$\langle\Phi_u^0(\mu)\rangle=v_u(\mu)=v(\mu)\sin\beta(\mu)$$
$$\langle\Phi_d^0(\mu)\rangle=v_d(\mu)=v(\mu)\cos\beta(\mu)$$
$$v^2(\mu)=v^2_u(\mu)+v^2_d(\mu)$$
\be\tan\beta(\mu)=v_u(\mu)/v_d(\mu)\label{eq3}\ee
The relevant RGEs for the Yukawa matrices at one-loop level for the three
effective theories are expressed as \cite{15,16,17,18,25,26,27},
\widetext
\ba\dudt&=&\left[{\rm Tr}\left(3\UU+3a\DD+a\EE\right)+{3\over 2}\left(b\UU+
c\DD\right)-\sum_iC_i^{(u)}g_i^2\right]Y_U\nonumber\\
\dddt&=&\left[{\rm Tr}\left(3a\UU+3\DD+\EE\right)+{3\over 2}\left(b\DD+
c\UU\right)-\sum_iC_i^{(d)}g_i^2\right]Y_D\nonumber\\
\dedt&=&\left[{\rm Tr}\left(3a\UU+3\DD+\EE\right)+{3\over 2}b\EE-\sum_i
C_i^{(e)}g_i^2\right]Y_E\label{eq4}\ea
\narrowtext
The RGEs for the VEV in the SM has been derived up to two-loop from wave
function renormalisation of the scalar field
\cite{15,16,18,19,21,22} and the one-loop equation is,
\mediumtext
\be\dvdt=\left[\sum C_i^vg_i^2-{\rm Tr}\left(3\UU+3\DD+\EE\right)
\right]v\label{eq5}\ee
where $t=\ln\mu$.
\narrowtext
\par
The RGEs for $v_a(a=u,d)$ in the 2HDM up to one-loop and in MSSM upto
two-loops have been derived in \cite{15,16,17,18,20}. The one-loop equations
in both the theories are,
\ba\dvudt&=&\left[\sum C_i^vg_i^2-{\rm Tr}\left(3\UU\right)\right]v_u
\nonumber\\
\dvddt&=&\left[\sum C_i^vg_i^2-{\rm Tr}\left(3\DD+\EE\right)\right]v_d
\label{eq6}\ea
The gauge couplings in the three models obey the well known one-loop RGEs,
\be\dgdt=b_ig_i^3\label{eq7}\ee
Two-loop contributions have been derived by a number of authors
\cite{15,16,17,18,21,22,23,24,25,26,27}.
The coefficients appearing in the RHS of (\ref{eq4})-(\ref{eq7}) are defined
in the three different cases,
\newpage\noindent
\underline{SM, 2HDM}
\ba C_i^u&=&\left({17\over 20},{9\over 4},8\right)\nonumber\\
C_i^d&=&\left({1\over 4},{9\over 4},8\right)\nonumber\\
C_i^e&=&\left({9\over 4},{9\over 4},0\right)\nonumber\\
C_i^v&=&\left({9\over 20},{9\over 4},0\right)\label{eq8}\ea
\underline{MSSM}
\ba C_i^u&=&\left({13\over 5},3,{16\over 3}\right)\nonumber\\
C_i^d&=&\left({7\over 15},3,{16\over 3}\right)\nonumber\\
C_i^e&=&\left({9\over 5},3,0\right)\nonumber\\
C_i^v&=&\left({3\over 20},{3\over 4},0\right)\label{eq9}\ea
\underline{SM}
\ba b_i&=&\left({41\over 10},-{19\over 6},-7\right)\nonumber\\
\left(a,b,c\right)&=&\left(1,1,-1\right)\label{eq10}\ea
\underline{2HDM}
\ba b_i&=&\left({21\over 5},-3,-7\right)\nonumber\\
\left(a,b,c\right)&=&\left(0,1,{1\over 3}\right)\label{eq11}\ea
\underline{MSSM}
\ba b_i&=&\left({33\over 5},1,-3\right)\nonumber\\
\left(a,b,c\right)&=&\left(0,2,{2\over 3}\right)\label{eq12}\ea
For the sake of simplicity we have neglected the Yukawa interactions of
neutrinos. Assuming that the right-handed neutrinos are massive
$(M_N>10^{13}$ GeV) our formulas are valid below $M_N$ to a very good
approximations even if such interactions are included.
\par
\section{RGEs and Analytic Formulas for Running Masses}
\label{sec3}
\par
Using the definition (\ref{eq1}) and (\ref{eq4})-(\ref{eq12}), we obtain the
RGEs for the mass matrices in the broken phases of SM, 2HDM, or MSSM in the
following form,
\ba\dMudt&=&\left(-\sum_iC_ig_i^2+\tilde a\UU+\tilde b\DD\right)M_U
\nonumber\\
\dMddt&=&\left(-\sum_iC'_ig_i^2+\tilde b\UU+\tilde a\DD\right)M_D
\nonumber\\
\dMedt&=&\left(-\sum_iC''_ig_i^2+\tilde c\EE\right)M_E\label{eq13}\ea
where the coefficients in the RHS are defined for the three cases.
\\
\underline{SM, 2HDM}
\ba C_i&=&\left({2\over 5},0,8\right)\nonumber\\
C_i'&=&\left(-{1\over 5},0,8\right)\nonumber\\
C_i''&=&\left({9\over 5},0,0\right)\label{eq14}\ea
\underline{MSSM}
\ba C_i&=&\left({49\over 20},{9\over 4},{16\over 3}\right)\nonumber\\
C_i'&=&\left({19\over 60},{9\over 4},{16\over 3}\right)\nonumber\\
C_i''&=&\left({33\over 20},{9\over 4},0\right)\nonumber\\
\left(\tilde a,\tilde b,\tilde c\right)&=&\left(3,1,3\right)\label{eq15}\ea
\underline{SM}
\be\left(\tilde a,\tilde b,\tilde c\right)=\left({3\over 2},-{3\over 2},
{3\over 2}\right)\label{eq16}\ee
\underline{2HDM}
\be\left(\tilde a,\tilde b,\tilde c\right)=\left({3\over 2},{1\over 2},
{3\over 2}\right)\label{eq17}\ee
Defining the diagonal mass matrices $\hat{M}_F$, the diagonal Yukawa matrices
$(\hat{Y}_F)$ and the CKM matrix $(V)$ through biunitary transformations
$L_F$ and $R_F$ on the left(right)-handed fermion $F_L(F_R)$ with $F=U,D,E$
\ba\hat{M}_F&=&L^\dagger_FM_FR_F\nonumber\\
\hat{Y}_F&=&L^\dagger_FY_FR_F\nonumber\\
\hat{M}^2_F&=&L^\dagger_FM_FM_F^\dagger L_F\nonumber\\
\hat{Y}^2_F&=&L^\dagger_FY_FY_F^\dagger L_F\nonumber\\
V&=&L^\dagger_UL_D\label{eq18}\ea
and following the procedures outlined in \cite{13,28} we obtain
\widetext
\ba\dMus&=&\left[\Mus,\uudot\right]+{1\over 16\pi^2}\left[-2\sum_iC_ig^2_i
\Mus+2\tilde a\hat{Y}^2_U\Mus+\tilde b\left(V\hat{Y}^2_DV^\dagger\Mus+\Mus V
\hat{Y}^2_DV^\dagger\right)\right]\nonumber\\
\dMds&=&\left[\Mds,\dddot\right]+{1\over 16\pi^2}\left[-2\sum_iC'_ig^2_i\Mds
+2\tilde a\hat{Y}^2_D\Mds+\tilde b\left(V^\dagger\hat{Y}^2_UV\Mds+\Mds V^
\dagger\hat{Y}^2_UV\right)\right]\nonumber\\
\dMes&=&\left[\Mes,\eedot\right]+{1\over 16\pi^2}\left[-2\sum_iC''_ig^2_i\Mes
+2\tilde c\hat{Y}^2_E\Mes\right]\label{eq19}\ea
\narrowtext
where $\dot{L}_F={dL_F\over dt}$.
\par
We point out that in the corresponding RGEs
for Yukawa couplings given by eq.~(2.13) in Ref.~\cite{28}, the terms
$-2\sum_iC^u_ig_i^2\hat{Y}^2_U/\left(16\pi^2\right)$,
$-2\sum_iC^d_ig_i^2\hat{Y}^2_D/\left(16\pi^2\right)$ and
$-2\sum_iC^e_ig_i^2\hat{Y}^2_E/\left(16\pi^2\right)$
are missing from the R.H.S.
\par
The diagonal elements of $L_F^\dagger\dot{L}_F(F=U,D,E)$ are made to vanish
in the usual manner by diagonal phase multiplication. The nondiagonal elements
of both sides of (\ref{eq19}) give the same RGEs for CKM matrix elements as
before which on integration yields \cite{28,29},
\mediumtext
\ba\left|V_{\alpha\beta}(\mu)\right|=\left\{\br{ll}\left|V_{\alpha\beta}(m_t)
\right|e^{-{3\over 2}c\left(I_t(\mu)+I_b(\mu)\right)},&\mbox{$\alpha\beta=ub,
cb, tb, ts$}\\
\left|V_{\alpha\beta}(m_t)\right|,& \mbox{otherwise}\er\right.\label{eq20}\ea
\narrowtext
Taking diagonal elements of both sides of (\ref{eq19}) and using dominance of
Yukawa couplings of the third generation over the first two, except the charm
quark, we obtain RGEs for the mass eigen values of quarks and leptons,
\ba\dmudt&=&\left[-\sum_iC_ig_i^2+\tilde by^2_b\left|V_{ub}\right|^2\right]
m_u\nonumber\\
\dmcdt&=&\left[-\sum_iC_ig_i^2+\tilde ay^2_c+\tilde by_b^2\left|V_{cb}\right|
^2\right]m_c\nonumber\\
\dmtdt&=&\left[-\sum_iC_ig_i^2+\tilde ay^2_t+\tilde by_b^2\left|V_{tb}\right|
^2\right]m_t\nonumber\\
\dmjdt&=&\left[-\sum_iC'_ig_i^2+\tilde by^2_t\left|V_{tj}\right|^2\right]m_j,
j=d,s\nonumber\\
\dmbdt&=&\left[-\sum_iC'_ig_i^2+\tilde ay^2_b+\tilde by_t^2\left|V_{tb}\right|
^2\right]m_b\nonumber\\
\dmjdt&=&\left[-\sum_iC''_ig_i^2\right]m_j,j=e,\mu\nonumber\\
\dmtaudt&=&\left[-\sum_iC''_ig_i^2+\tilde cy^2_{\tau}\right]m_{\tau}
\label{eq21}\ea
Integrating (\ref{eq21}) and using the corresponding low-energy values, the
new analytic formulas are obtained in the small mixing limit as
\ba m_u(\mu)&=&m_u(1~{\rm GeV})\eta_u^{-1}B_u^{-1}\nonumber\\
m_c(\mu)&=&m_c(m_c)\eta_c^{-1}B_u^{-1}e^{\tilde a I_c}\nonumber\\
m_t(\mu)&=&m_t(m_t)B_u^{-1}e^{\tilde aI_t+\tilde bI_b}\nonumber\\
m_i(\mu)&=&m_i(1~{\rm GeV})\eta_i^{-1}B_d^{-1},i=d,s\nonumber\\
m_b(\mu)&=&m_b(m_b)\eta_b^{-1}B_d^{-1}e^{\tilde aI_b+\tilde bI_t}\nonumber\\
m_i(\mu)&=&m_i(1~{\rm GeV})\eta_i^{-1}B_e^{-1},i=e,\mu\nonumber\\
m_\tau(\mu)&=&m_\tau(m_\tau)\eta_\tau^{-1}B_e^{-1}e^{\tilde cI_\tau}
\label{eq22}\ea
where
\ba B_u&=&\prod\left({\alpha_i(\mu)\over\alpha_i(m_t)}\right)^{C_i\over 2b_i}
\nonumber\\
B_d&=&\prod\left({\alpha_i(\mu)\over\alpha_i(m_t)}\right)^{C'_i\over 2b_i}
\nonumber\\
B_e&=&\prod\left({\alpha_i(\mu)\over\alpha_i(m_t)}\right)^{C''_i\over 2b_i}
\label{eq23}\ea
\be I_f(\mu)={1\over 16\pi^2}\int_{\ln m_t}^{\ln\mu}y^2_f(t')dt'
\label{eq24}\ee
The ratio $\eta_f (f=u, d, c, s, b, e, \mu, \tau)$ appearing in (\ref{eq22})
is the QCD-QED rescaling factor for fermion mass $m_f$.
\par
Comparison with earlier and recent derivation of analytic formulas
\cite{30,31} shows several differences. Whereas the top-quark Yukawa coupling
integral in (\ref{eq24}) has been predicted to affect the running of
$m_u(\mu)$ and $m_c(\mu)$ our formulas predict no such dependence. Similarly,
whereas the $b$-quark and the $\tau$-lepton Yukawa coupling integrals have
been predicted to affect the running charged-lepton masses $m_e(\mu)$ and
$m_\mu(\mu)$, our formulas predict no such contributions. Our formulas predict
that in all the three cases, SM, 2HDM, or MSSM, the third generation Yukawa
couplings do not affect the running masses of the first two generations in the
small mixing limit. Also for the third generation running masses, the Yukawa
coupling integrals occur in the exponential factors with different
coefficients as compared to \cite{30,31}. The dependence on gauge couplings
are also noted to be quite different in our analytic formulas. Whereas earlier
derivation \cite{30} predicted the occurrence of the exponents $C^u_i/2b_i$,
$C^d_i/2b_i$, $C^e_i/2b_i$ on the RHS of (\ref{eq23}), our formulas predict
the corresponding exponents to be $C_i/2b_i$, $C'_i/2b_i$, $C''_i/2b_i$
respectively. Thus, the new analytic formulas derived here at one-loop level
predict substantially new dependence on gauge and Yukawa couplings for the
running masses. Our formulas for the case of MSSM are the same as those
obtained in \cite{13} where the SUSY scale was taken as $M_S=m_t$. As derived
in \cite{13} for the MSSM, the formula for running $\tan\beta(\mu)$ has also
the same form also in the 2HDM at one-loop level.
\mediumtext
\be\tan\beta(\mu)=\tan\beta(m_t)\exp\left[-3I_t(\mu)+3I_b(\mu)+I_\tau(\mu)
\right]\label{eq25}\ee
\narrowtext
In contrast, when running of VEVs are ignored, apart from the mass predictions
being different, $\tan\beta$ is also predicted to be the same for all values
of $\mu>M_Z$ in 2HDM or MSSM \cite{30,31}.
\par
\section{Formulas in MSSM for $M_S>M_Z$}
\label{sec4}
\par
In the MSSM the natural SUSY scale $(M_S)$ could be very different from the
weak scale with $M_S\approx O$ (TeV), whereas $M_S\gg 1$ TeV has gauge
hierarchy problem. As our new contribution in MSSM in this paper, compared to
\cite{13}, we present new analytic formulas for all charged fermion masses for
any SUSY scale $M_S>M_Z$ by running them from $m_t-M_S$ as in SM and then from
$M_S-\mu$ as in MSSM.
\mediumtext
\ba m_u(\mu)&=&m_u(1~{\rm GeV})\eta_u^{-1}G_u(\mu)\label{eq26}\\
m_c(\mu)&=&m_c(m_c)\eta_c^{-1}G_u(\mu)\exp\left({3\over 2}I_c(M_S)+3\tilde
I_c(\mu)\right)\label{eq27}\\
m_t(\mu)&=&m_t(m_t)G_u(\mu)\exp\left({3\over 2}I_t(M_S)-{3\over 2}I_b(M_S)+
3\tilde I_t(\mu)+\tilde I_b(\mu)\right)\label{eq28}\\
m_i(\mu)&=&m_i(1~{\rm GeV})\eta_i^{-1}G_d(\mu), i=d,s\label{eq29}\\
m_b(\mu)&=&m_b(m_b)\eta_b^{-1}G_d(\mu)\exp\left({3\over 2}I_b(M_S)-
{3\over 2}I_t(M_S)+3\tilde I_b(\mu)+\tilde I_t(\mu)\right)\label{eq30}\\
m_i(\mu)&=&m_i(1~{\rm GeV})\eta_i^{-1}G_e(\mu), i=e,\mu\label{eq31}\\
m_\tau(\mu)&=&m_\tau(m_\tau)\eta_\tau^{-1}G_e(\mu)\exp\left({3\over 2}I_
\tau(M_S)+3\tilde I_\tau(\mu)\right)\label{eq32}\ea
where
\ba
G_u(\mu)&=&\left({\alpha_1(M_S)\over\alpha_1(m_t)}\right)^{-2\over 41}
\left({\alpha_3(M_S)\over\alpha_3(m_t)}\right)^{4\over 7}
\left({\alpha_1(\mu)\over\alpha_1(M_S)}\right)^{-49\over 264}
\left({\alpha_2(\mu)\over\alpha_2(M_S)}\right)^{-9\over 8}
\left({\alpha_3(\mu)\over\alpha_3(M_S)}\right)^{8\over 9}
\nonumber\\
G_d(\mu)&=&\left({\alpha_1(M_S)\over\alpha_1(m_t)}\right)^{1\over 41}
\left({\alpha_3(M_S)\over\alpha_3(m_t)}\right)^{4\over 7}
\left({\alpha_1(\mu)\over\alpha_1(M_S)}\right)^{-19\over 792}
\left({\alpha_2(\mu)\over\alpha_2(M_S)}\right)^{-9\over 8}
\left({\alpha_3(\mu)\over\alpha_3(M_S)}\right)^{8\over 9}
\nonumber\\
G_e(\mu)&=&\left({\alpha_1(M_S)\over\alpha_1(m_t)}\right)^{-9\over 41}
\left({\alpha_1(\mu)\over\alpha_1(M_S)}\right)^{-1\over 8}
\left({\alpha_2(\mu)\over\alpha_2(M_S)}\right)^{-9\over 8}
\label{eq33}\ea
\narrowtext
\be\tilde I_f(\mu)={1\over 16\pi^2}\int_{\ln M_S}^{\ln\mu}y^2_f(t')dt'
\label{eq34}\ee
and $I_f(M_S)$ is defined through (\ref{eq24}) with $\mu=M_S$. Running of the
elements of the CKM matrix in the MSSM are modified by the following formulas
\mediumtext
\ba\left|V_{\alpha\beta}(\mu)\right|=\left\{\br{ll}\left|V_{\alpha\beta}(m_t)
\right|e^{{3\over 2}\left(I_t(M_S)
+I_b(M_S)\right)}e^{-\left(\tilde I_t(\mu)+\tilde I_b(\mu)\right)},&
\mbox{$\alpha\beta=ub,cb,tb,ts$}\\
\left|V_{\alpha\beta}(m_t)\right|,&\mbox{otherwise}\er\right.\label{eq35}\ea
\narrowtext
The one-loop formula for $\tan\beta(\mu)$ in (\ref{eq25}) is also modified,
\mediumtext
\be\tan\beta(\mu)=\tan\beta(M_S)\exp\left(-3\tilde I_t(\mu)+3\tilde I_b(\mu)
+\tilde I_\tau(\mu)\right)\label{eq36}\ee
\narrowtext
The analytic formulas (\ref{eq26})-(\ref{eq37}) hold good for any value of
$m_t<M_S<\mu$. It may be noted that in the limit of $M_S\to m_t$,
$I_f(M_S)\to 0$, $\tilde I_f(\mu)\to I_f(\mu)$ and the formulas
(\ref{eq26})-(\ref{eq36}) reduce to those obtained in \cite{13}. It is to be
noted that corresponding exponent in the expression $B_u$ in eq.~(3.6) of
ref.~[13] should be corrected as ${49\over 264}$ in place of ${43\over 792}$.
\par
\section{Numerical Predictions at Higher Scales}
\label{sec5}
\par
The analytic formulas given in the previous section predict masses and the CKM
matrix elements upto one-loop level at higher scales. We have also estimated
numerically the effect of scale dependent VEVs on predictions of the running
masses at two-loop level. We solve RGEs for the Yukawa matrices and VEVs
including two loop contributions in SM and MSSM \cite{15,16,17,18,25,26,27}
numerically and obtain the mass matrices at higher scales from the
corresponding products of the two. For this purpose the elements of the CKM
matrix at higher scales have been obtained by running them through one-loop
RGEs given by (\ref{eq20}) with appropriate values of the coefficient $c$ given
in (\ref{eq10})-(\ref{eq12}) \cite{28,29}. In 2HDM we carry out all numerical
estimations at one-loop level. We use the following inputs for the running
masses ($m_i$), SM gauge couplings ($\alpha_1$,$\alpha_2$,$\alpha_3$),
electromagnetic finestructure constant($\alpha$), electroweak mixing angle and
the CKM matrix $(V)$ at $\mu=M_Z$ which have been obtained from the
experimental data \cite{32}.
\ba m_u&=&2.33^{+0.42}_{-0.45}~{\rm MeV},m_c=677^{+56}_{-61}~{\rm MeV},
\nonumber\\
m_t&=&181\pm 13~{\rm GeV},m_d=4.69^{+0.60}_{-0.66}~{\rm MeV},\nonumber\\
m_s&=&93.4^{+11.8}_{-13.0}~{\rm MeV},m_b=3.00\pm 0.11~{\rm GeV},\nonumber\\
m_e&=&0.48684727\pm 0.00000014~{\rm MeV},\nonumber\\
m_\mu&=&102.75138\pm 0.00033~{\rm MeV},\nonumber\\
m_\tau&=&1.74669^{+0.00030}_{-0.00027}~{\rm GeV}\label{eq37}\ea
\ba\alpha_1(M_Z)&=&0.016829\pm 0.000017\nonumber\\
\alpha_2(M_Z)&=&0.033493^{+0.000042}_{-0.000038}\nonumber\\
\alpha_3(M_Z)&=&0.118\pm 0.003\nonumber\\
\alpha^{-1}_{em}&=&128.896\pm 0.09\nonumber\\
\sin^2\theta_W&=&0.23165\pm 0.000024\label{eq38}\ea
\mediumtext
\be V(M_Z)=\left(\br{lll}
0.9757,&0.2205,&0.0030e^{-i\delta}\\
-0.2203-0.0001e^{i\delta},&0.9747,&0.0373\\
0.0082-0.0029e^{i\delta},&-0.0364-0.0007e^{i\delta},&0.9993
\er\right)\label{eq39}\ee
\narrowtext
For the sake of convince we have used $\delta=\pi/2$ as in \cite{32}. The
choice of same inputs enables us to compare our results on mass predictions
with those obtained with scale-independent assumption on the VEVs in SM and
MSSM \cite{32}. We neglect mixings among charged-leptons and use the diagonal
basis for up-quarks.
\par
\begin{figure}
\epsfxsize=14cm
{\epsfbox{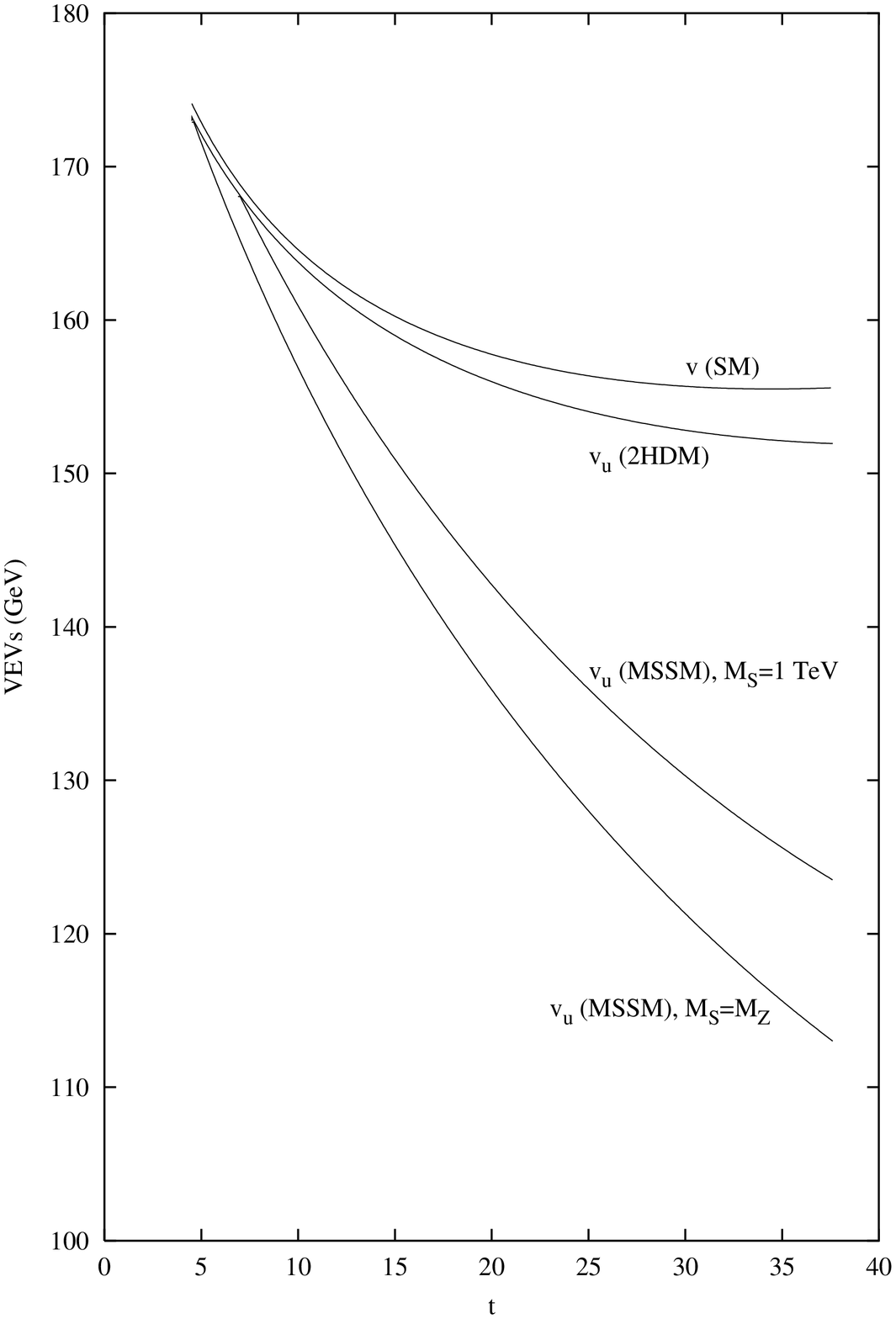}}
\caption{Variation of running VEVs in the SM, 2HDM and MSSM as a function of
$\mu (t=\ln\mu)$ showing substantial deviation from the scale-independent
assumption.}
\label{fig1}
\end{figure}
\begin{figure}
\epsfxsize=14cm
{\epsfbox{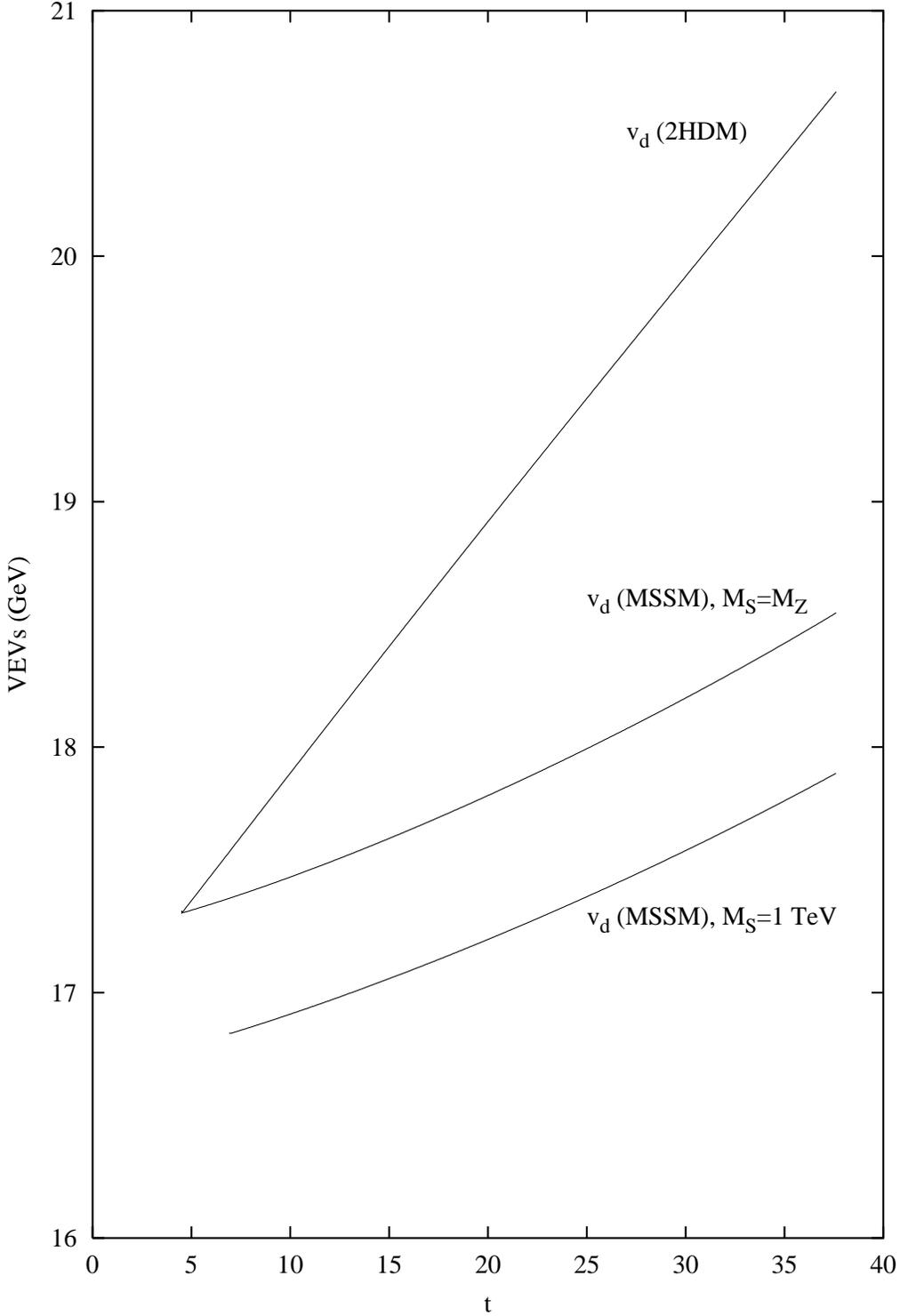}}
\caption{Variation of running VEVs at higher scales in MSSM and 2HDM as a
function of $\mu (t=\ln\mu)$ showing substantial deviation from the
scale-independent assumption.}
\label{fig2}
\end{figure}

The variations of VEVs as a function of $\mu$ are shown in
Figs.~\ref{fig1}-\ref{fig2} for the SM, 2HDM, and MSSM where the initial value
of $\tan\beta(M_Z)=10$ has been used for the latter two cases. In these and
certain other Figs.~we have used the variable $t=\ln\mu$ along X-axis where
$\mu$ is in units of GeV. It is quite clear that in the SM as well as the
other cases the running effects of the VEVs contribute to very significant
departures from the assumed scale-independent values \cite{28,30,31,32}. Thus
the predicted running masses are to be different in all the three cases. Since
$v_u(\mu)$ increases and $v_d(\mu)$ decreases with increasing  $\mu$, the
up-quark masses are expected to have decreasing effects whereas the down-quark
and charged-lepton masses are expected to have increasing effects at higher
scales in MSSM and 2HDM. But in the SM all the masses are expected to have
decreasing effects due to decreasing value of $v(\mu)$. In fact these features
are clearly exhibited in all numerical values of mass predictions carried out
in this investigation. It is to be noted that almost all fermion masses,
except the top-quark, $b$-quark and the $\tau$-lepton near the perturbative
limits, decrease at higher scales due to decrease in corresponding Yukawa
couplings. But the effect of running VEVs contribute to additional decreasing
or increasing factors in the respective cases.
\par
\begin{figure}
\epsfxsize=14cm
{\epsfbox{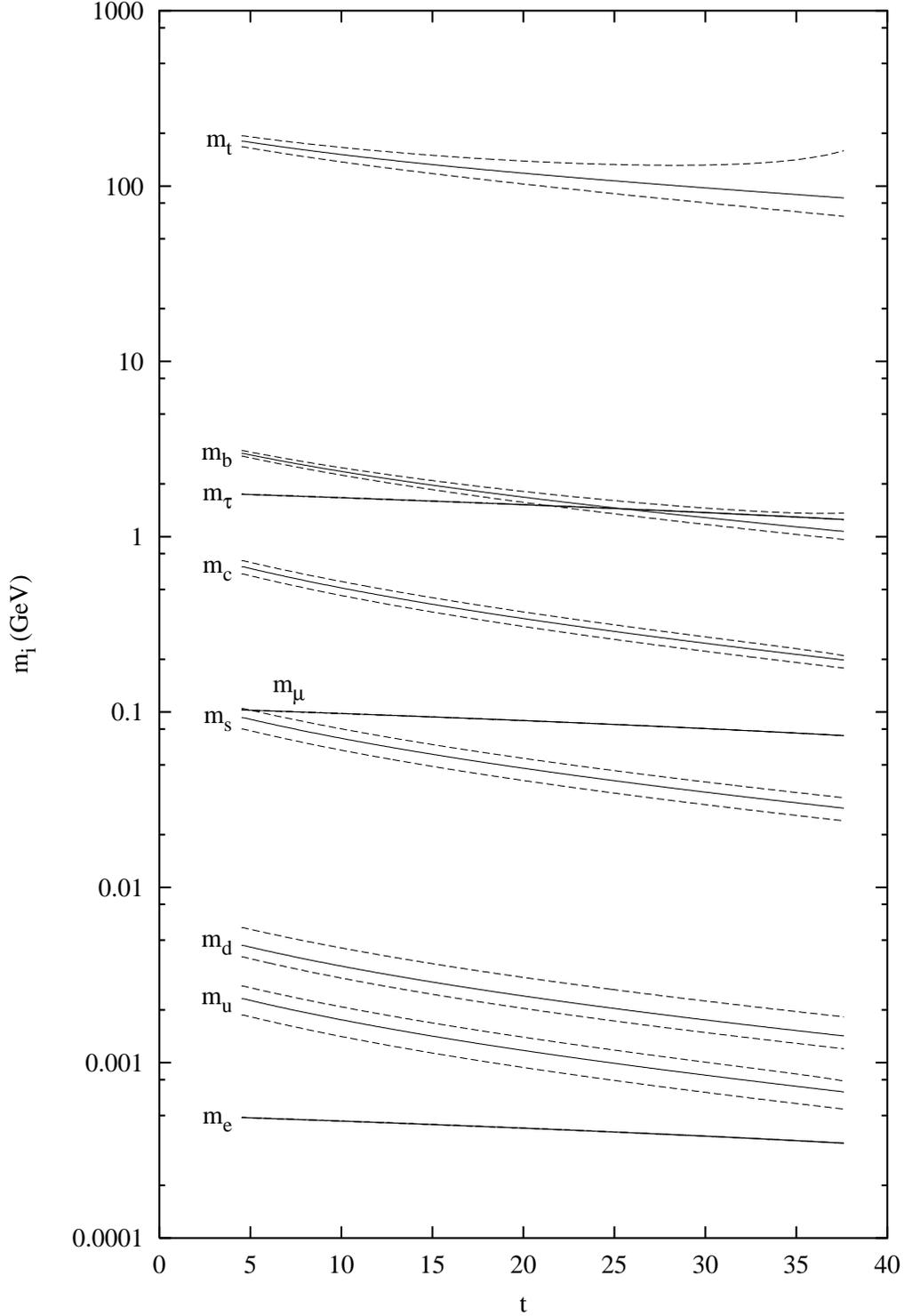}}
\caption{Predictions of running masses at higher scales as a function of $\mu$
($t=ln\mu$) in MSSM with SUSY scale $M_S=M_Z$ using the input parameters given
in (\ref{eq37})-(\ref{eq39}) and $\tan\beta(M_S)=10$. The dashed lines are due
to uncertainties in the input parameters.}
\label{fig3}
\end{figure}
\begin{figure}
\epsfxsize=14cm
{\epsfbox{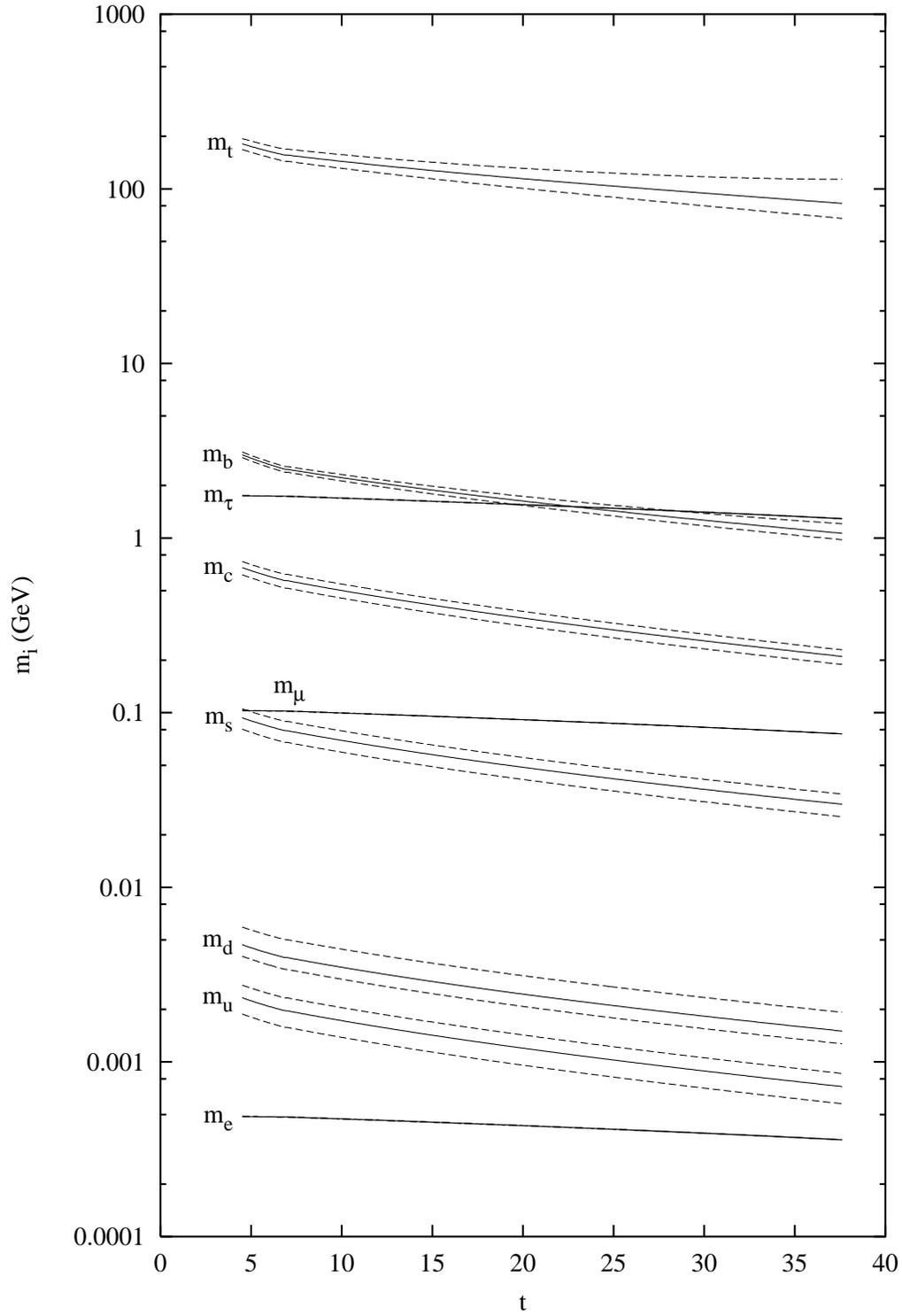}}
\caption{Same as Fig.~\ref{fig3} but with $M_S=1$ TeV.}
\label{fig4}
\end{figure}
\begin{figure}
\epsfxsize=14cm
{\epsfbox{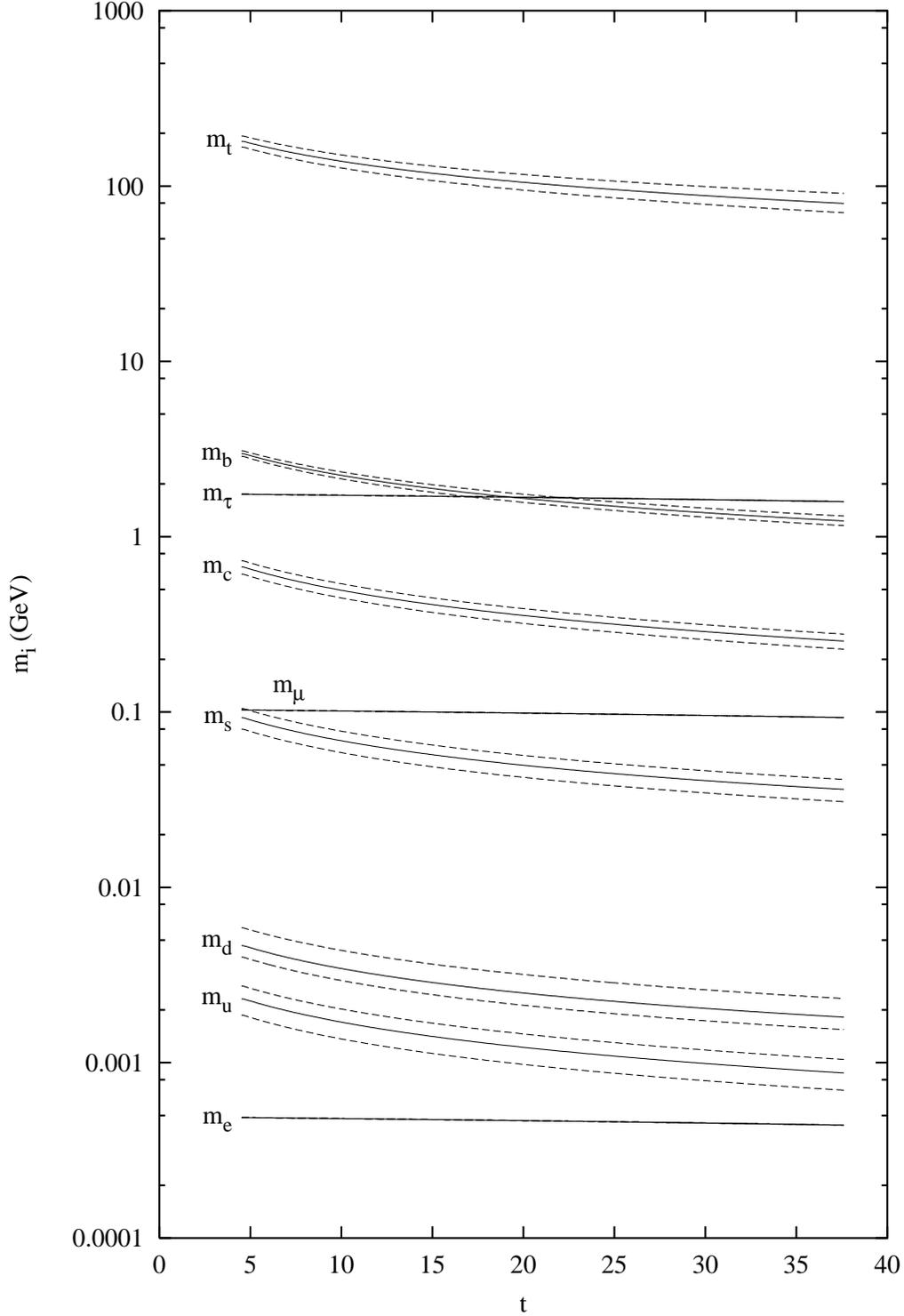}}
\caption{Predictions of running masses at higher scales in the 2HDM using the
input parameters given in (\ref{eq37})-(\ref{eq39}) and $\tan\beta(M_Z)=10$.}
\label{fig5}
\end{figure}
\begin{figure}
\epsfxsize=14cm
{\epsfbox{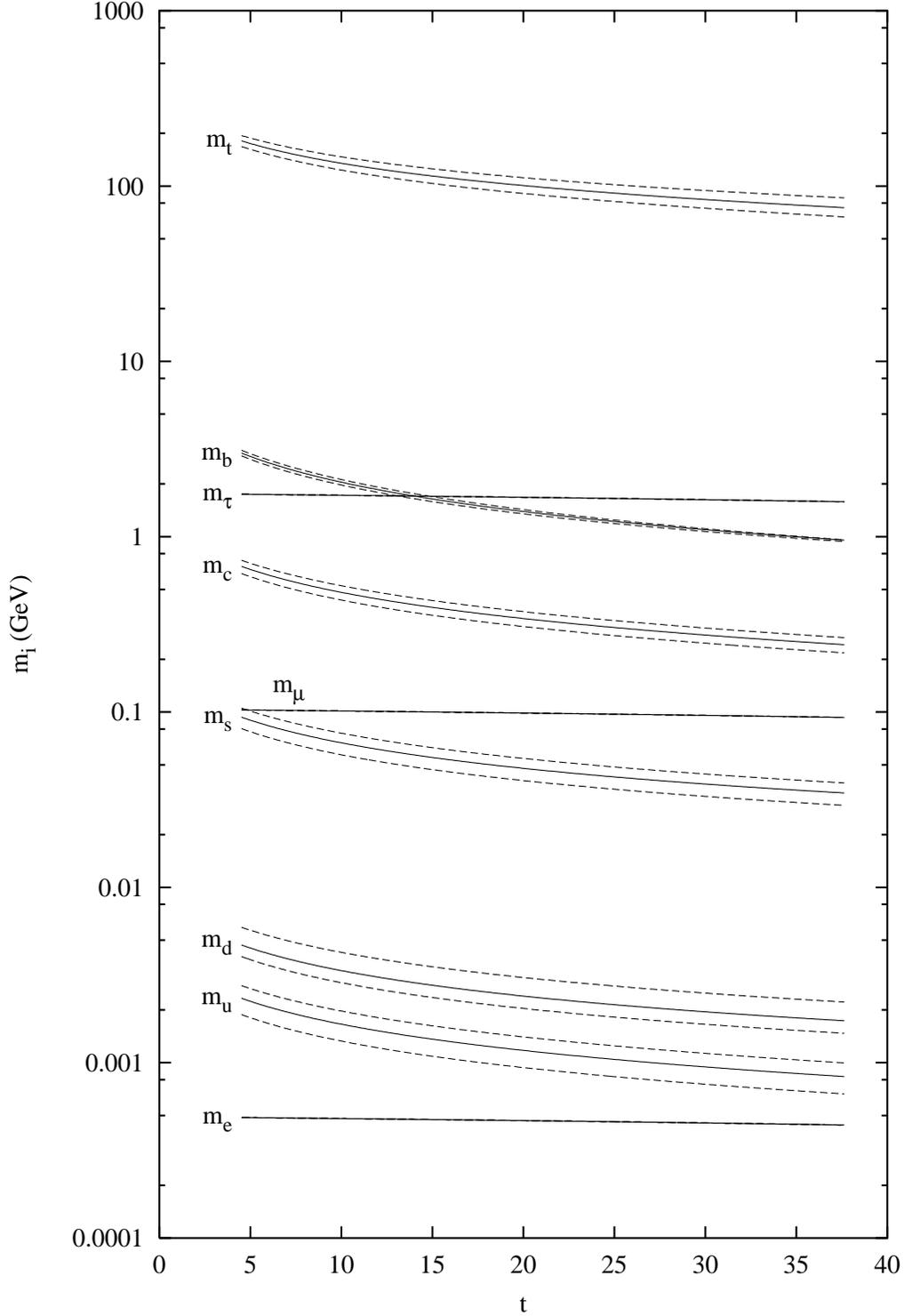}}
\caption{Predictions of running masses at higher scales in SM with the input
parameters given in (\ref{eq37})-(\ref{eq39}) and $M_H=250$ GeV.}
\label{fig6}
\end{figure}
\begin{figure}
\epsfxsize=14cm
{\epsfbox{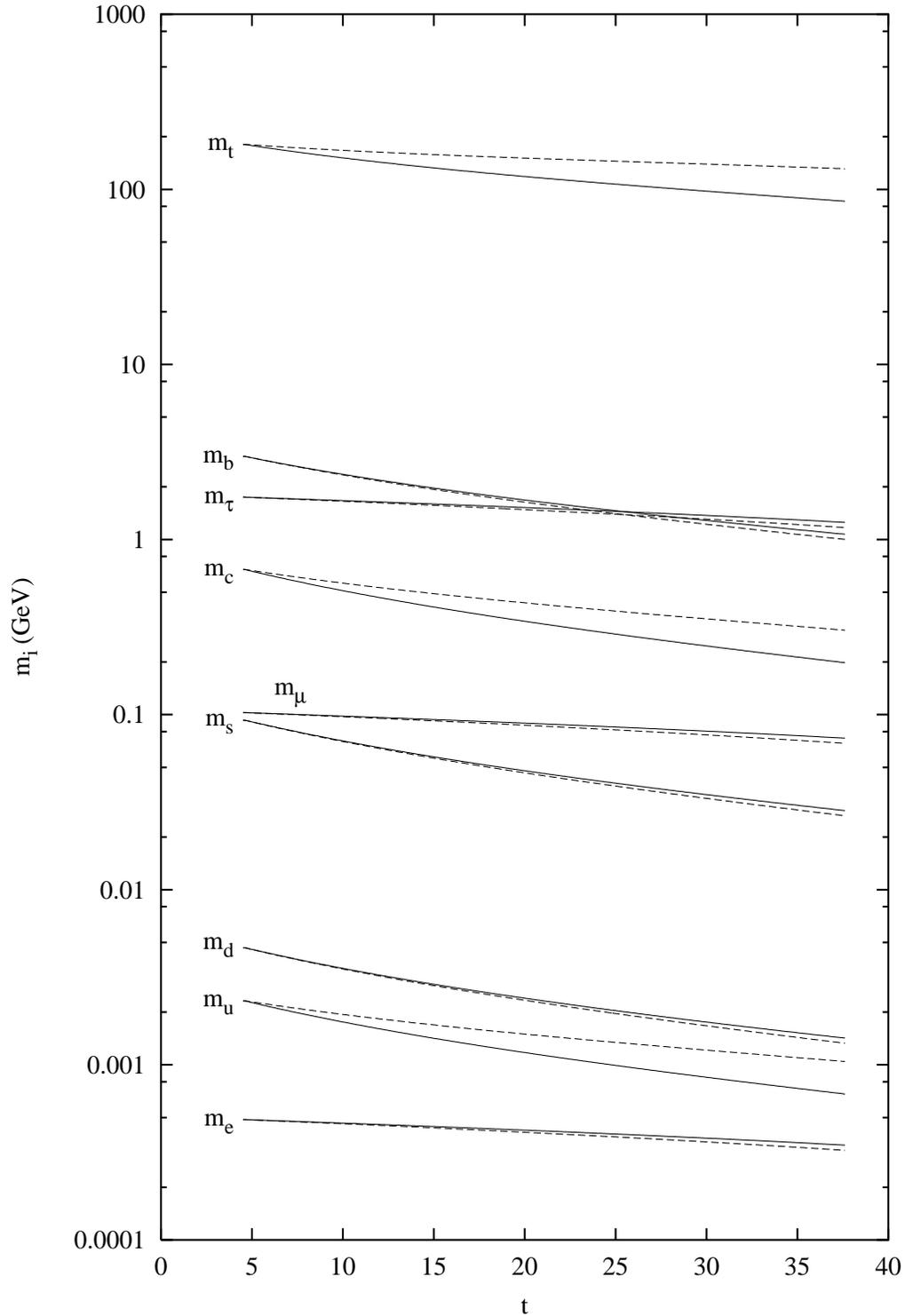}}
\caption{Comparison of running mass predictions in the MSSM (solid lines) with
those obtained from scale-independent assumptions (dashed lines) on the VEVs.
The SUSY scale has been taken to be $M_Z$.}
\label{fig7}
\end{figure}
\begin{figure}
\epsfxsize=14cm
{\epsfbox{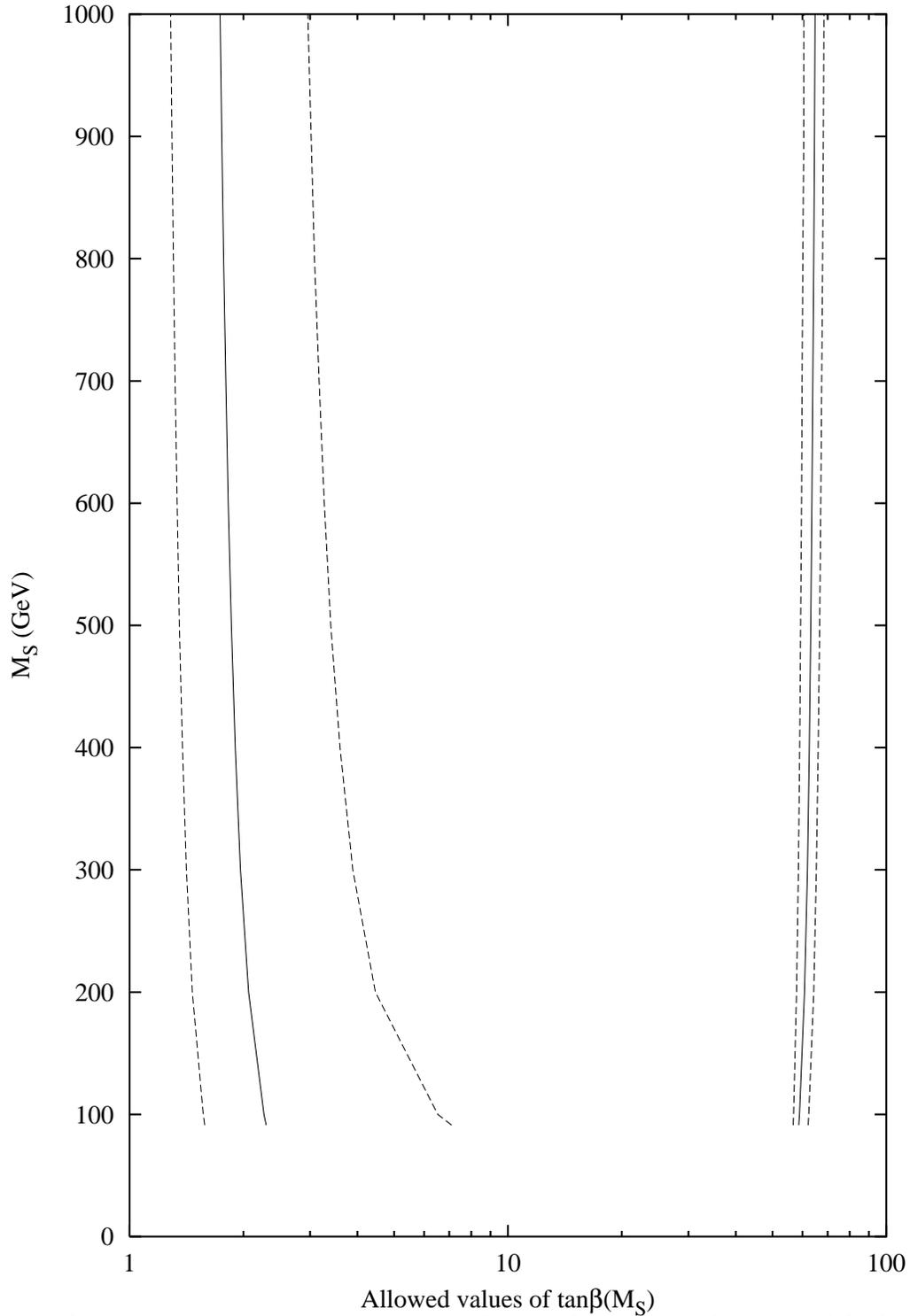}}
\caption{Perturbatively allowed region for $\tan\beta(M_S)$ as a function of
SUSY scale $M_S$. The lower (upper) limits are due to top-quark ($b$-quark)
Yukawa coupling. The dashed lines are due to uncertainties in the respective
input masses.}
\label{fig8}
\end{figure}

The predictions of all the charged fermion masses as a function of $t=\ln\mu$
are shown in Fig.~\ref{fig3} with $M_S=M_Z$ and in Fig.~\ref{fig4} with
$M_S=1$ TeV in the case of MSSM using $\tan\beta(M_Z)=10$. The corresponding
predictions in 2HDM and SM are shown in Figs.~\ref{fig5}-\ref{fig6}. In
Fig.~\ref{fig7} we display the comparison of mass predictions as functions of
$t=\ln\mu$ with and without running VEVs in MSSM assuming $M_S=M_Z$ and
$\tan\beta=10$. Although the differences in the two types of predictions are
clearly distinguishable, they are quite prominent in the up-quark sectors.
While the new contributions are seen to be significant for the down-quarks and
charged-leptons at higher scales with $\mu\ge 10^7$ GeV, in the case of
up-quarks the contributions are found to be important starting from $\mu=O$
(TeV). As compared to the scale-independent assumption \cite{32}, our
predictions are clearly smaller for
the up-quarks and larger for the down-quarks and charged-leptons as indicated
by solid-line curves in Fig.~\ref{fig7}. With the input values for $m_t$ and
$m_b$ in (\ref{eq37}), the lowest allowed value of $\tan\beta(M_S)$ is
determined by observing the perturbative limit for the top-quark Yukawa
coupling at the GUT scale, $y_t^2(M_{GUT})/{4\pi}\le 1.0$ and the highest
allowed value of $\tan\beta(M_S)$ is determined from the corresponding limit
on the $b$-quark Yukawa coupling.
\\
\underline{MSSM}
\ba M_S&=&M_Z:2.3^{+4.8}_{-0.6} \le\tan\beta(M_S)\le 58.7^{+3.4}_{-2.0},
\label{eq40}\\
M_S&=&1~{\rm TeV}:1.7^{+1.3}_{-0.4} \le\tan\beta(M_S)\le 64.8^{+3.6}_{-4.3}
\label{eq41}\ea
The allowed region for $\tan\beta(M_S)$ as a function of $M_S$ in MSSM is
shown in Fig.~\ref{fig8} where the solid (dashed) lines are due to the central
values (uncertainties) in the inputs of $m_t$ and $m_b$. It is clear that the
allowed region for $\tan\beta$ increases, although slowly, with increasing
$M_S$. In the 2HDM the allowed region for $\tan\beta$ is found to be
substantially larger.
\\                                                   
\underline{2HDM}
\be 1.2^{+0.3}_{-0.2} \le\tan\beta(M_Z) \le 68.9 \pm 2.7\label{eq42}\ee
\par
We have noted that in all the three effective theories the difference between
one- and two-loop estimation of running masses at the highest scale ($M_U$)
varies between 1-5\%, the lowest discrepancy being for the leptons and the
highest being for the top-quark. But in MSSM and 2HDM this discrepancy
increases  to 10-12\% for the $b$- and the top-quarks as the respective
perturbative limits are approached.
\par
\begin{figure}
\epsfxsize=14cm
{\epsfbox{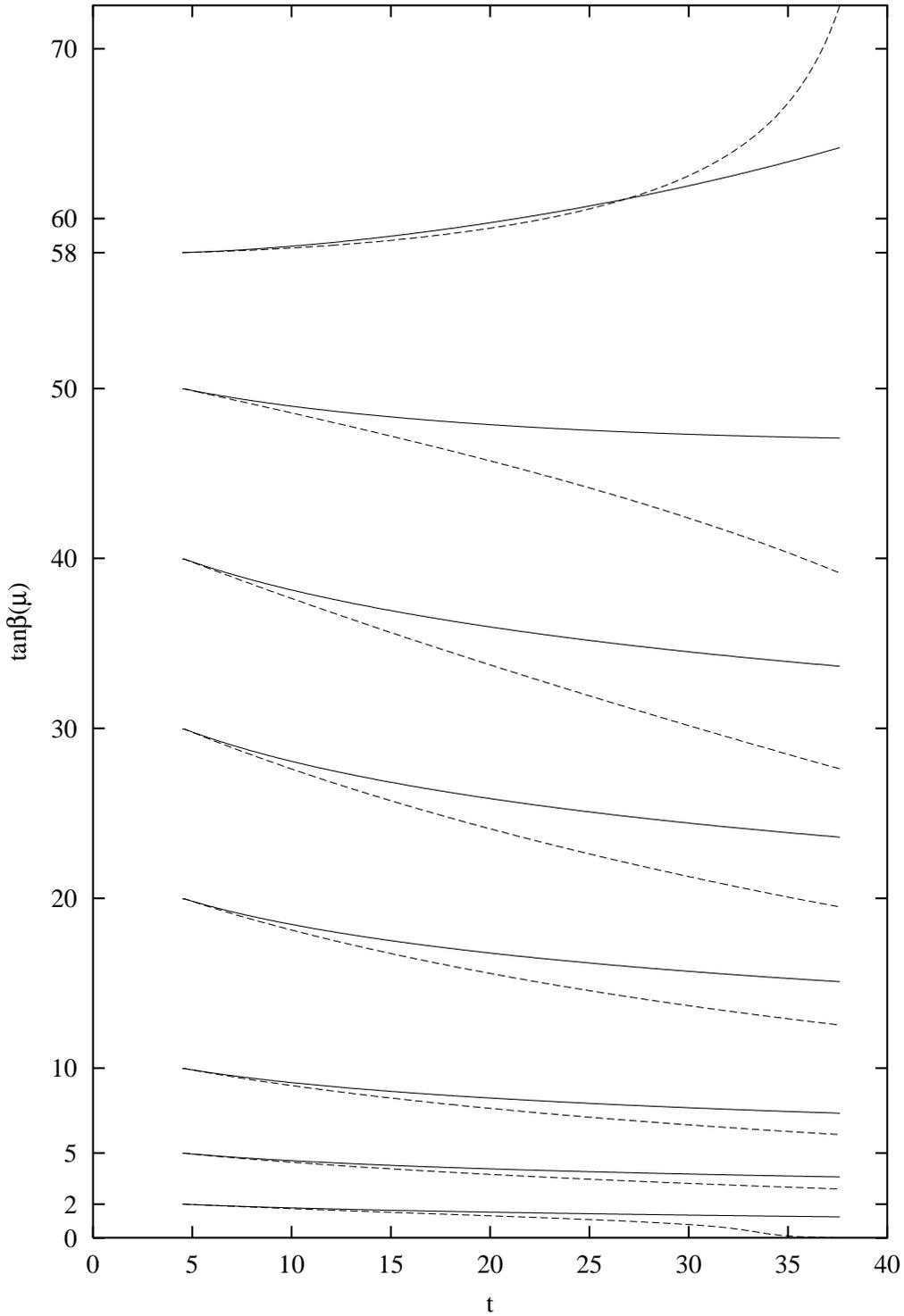}}
\caption{Variation of $\tan\beta(\mu)$ as a function of $\mu$ $(t=\ln\mu)$ for
different input values of $\tan\beta(M_Z)$ in MSSM (solid lines) and 2HDM
(dashed lines). The inputs for different curves starting from the bottom most
line is $\tan\beta(M_Z)=$2, 5, 10, 20, 30, 40, 50 and 58. In the MSSM the SUSY
scale has been taken as $M_Z$}
\label{fig9}
\end{figure}
\begin{figure}
\epsfxsize=14cm
{\epsfbox{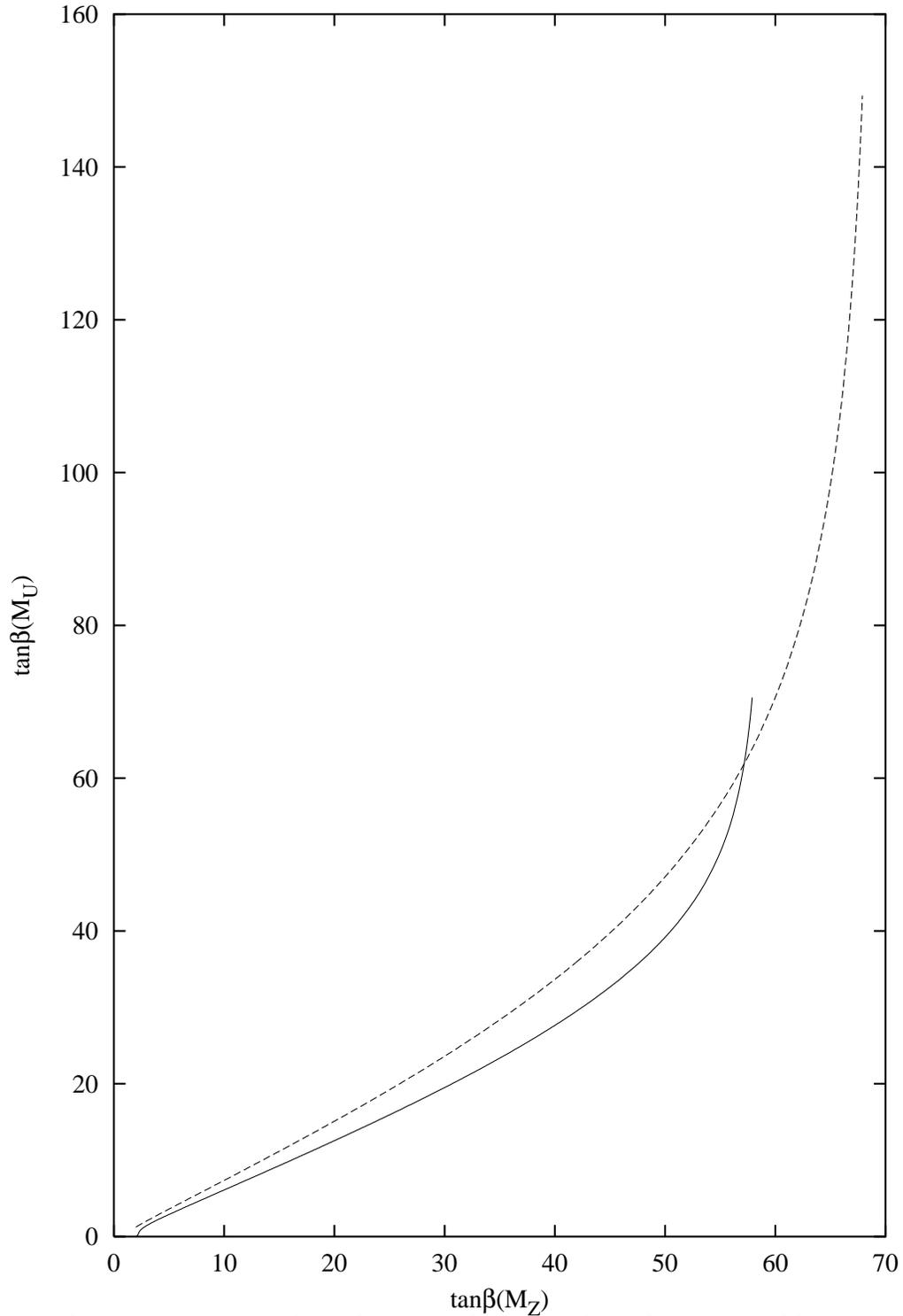}}
\caption{Predictions of $\tan\beta(M_U)$ as function of $\tan\beta(M_Z)$ in
MSSM (solid line) and 2HDM (dashed line). In MSSM the SUSY scale has been
taken as $M_Z$.}
\label{fig10}
\end{figure}

The running VEVs in MSSM and 2HDM lead to variation of $\tan{\beta}(\mu)$ as a
function of $\mu$ over its initial value at $M_Z$. This is shown in
Fig.~\ref{fig9} for different input values where the dashed (solid) line
represents the case for 2HDM (MSSM). In both the theories $\tan\beta(\mu)$
decreases (increases) from its initial value when the latter crosses a
critical point. This critical value is $\tan{\beta}(M_Z)\approx 56$ (52) in
MSSM (2HDM). In Fig.~\ref{fig10} we present $\tan{\beta}(M_U)$ at the GUT
scale as a function of $\tan{\beta}(M_Z)$ for both the theories. We observe
steep rise in the curves as the respective perturbative limits are approached
in the large $\tan\beta(M_Z)$ region.
\par
\begin{figure}
\epsfxsize=14cm
{\epsfbox{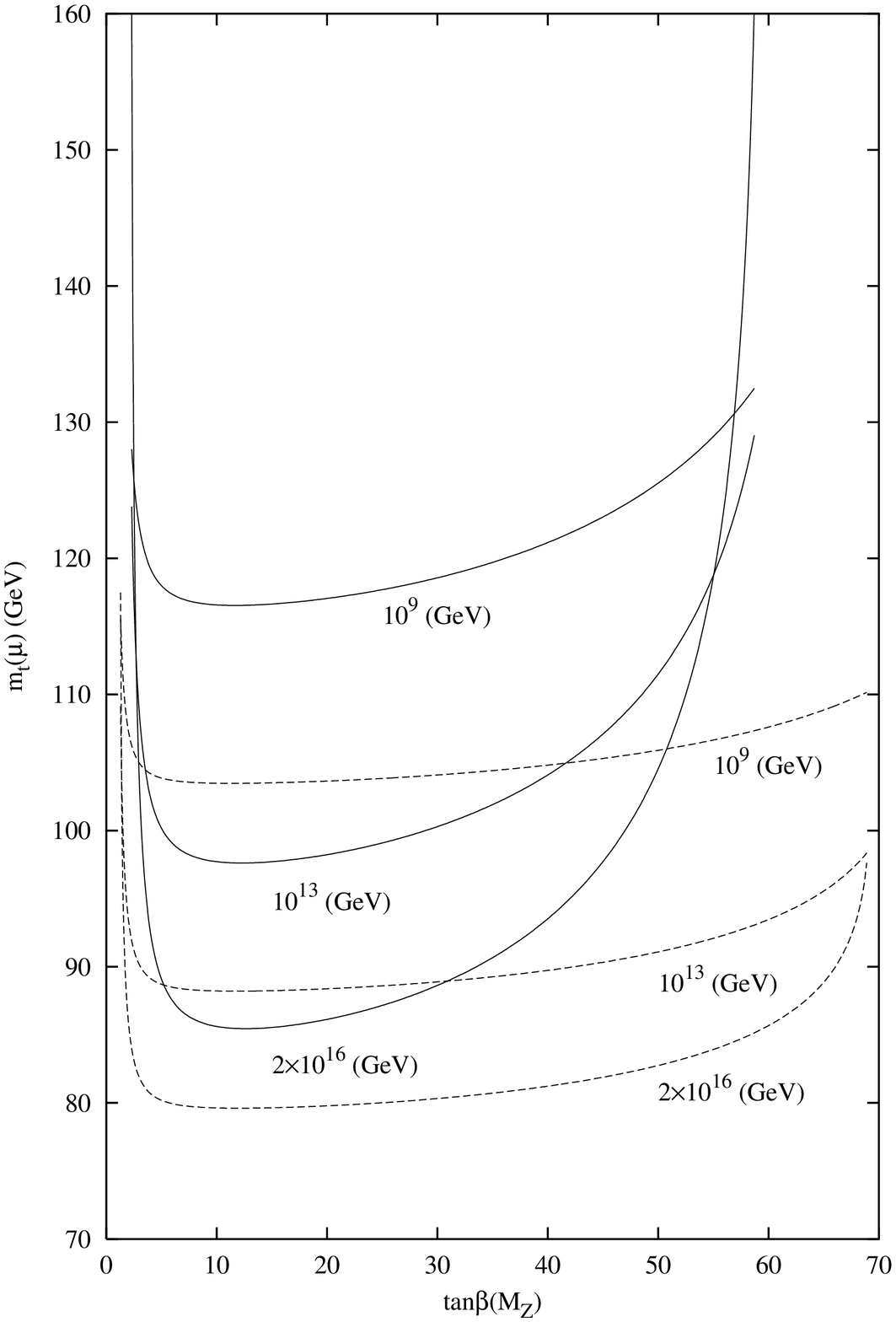}}
\caption{Prediction of $m_t(\mu)$ at higher scales, $\mu=10^9$ GeV, $10^{13}$
GeV and $2\times 10^{16}$ GeV as a function of $\tan\beta(M_Z)$ in MSSM with
$M_S=M_Z$ (solid lines) and 2HDM (dashed lines).}
\label{fig11}
\end{figure}
\begin{figure}
\epsfxsize=14cm
{\epsfbox{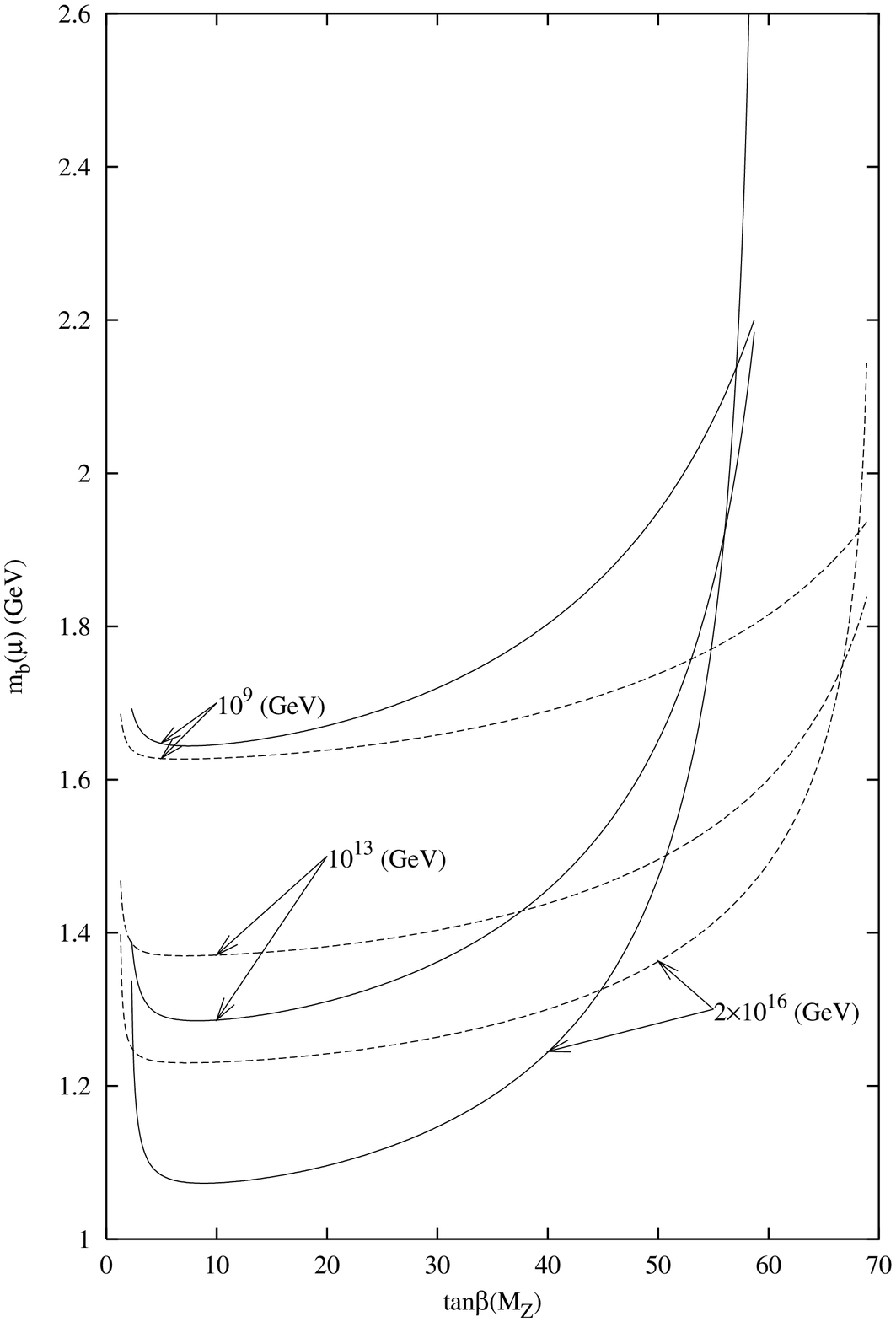}}
\caption{Predicton of $m_b(\mu)$ at higher scales, $\mu=10^9$ GeV, $10^{13}$
GeV and $2\times 10^{16}$ GeV as a function of $\tan\beta(M_Z)$ in MSSM with
$M_S=M_Z$ (solid lines) and 2HDM (dashed lines).}
\label{fig12}
\end{figure}
\begin{figure}
\epsfxsize=14cm
{\epsfbox{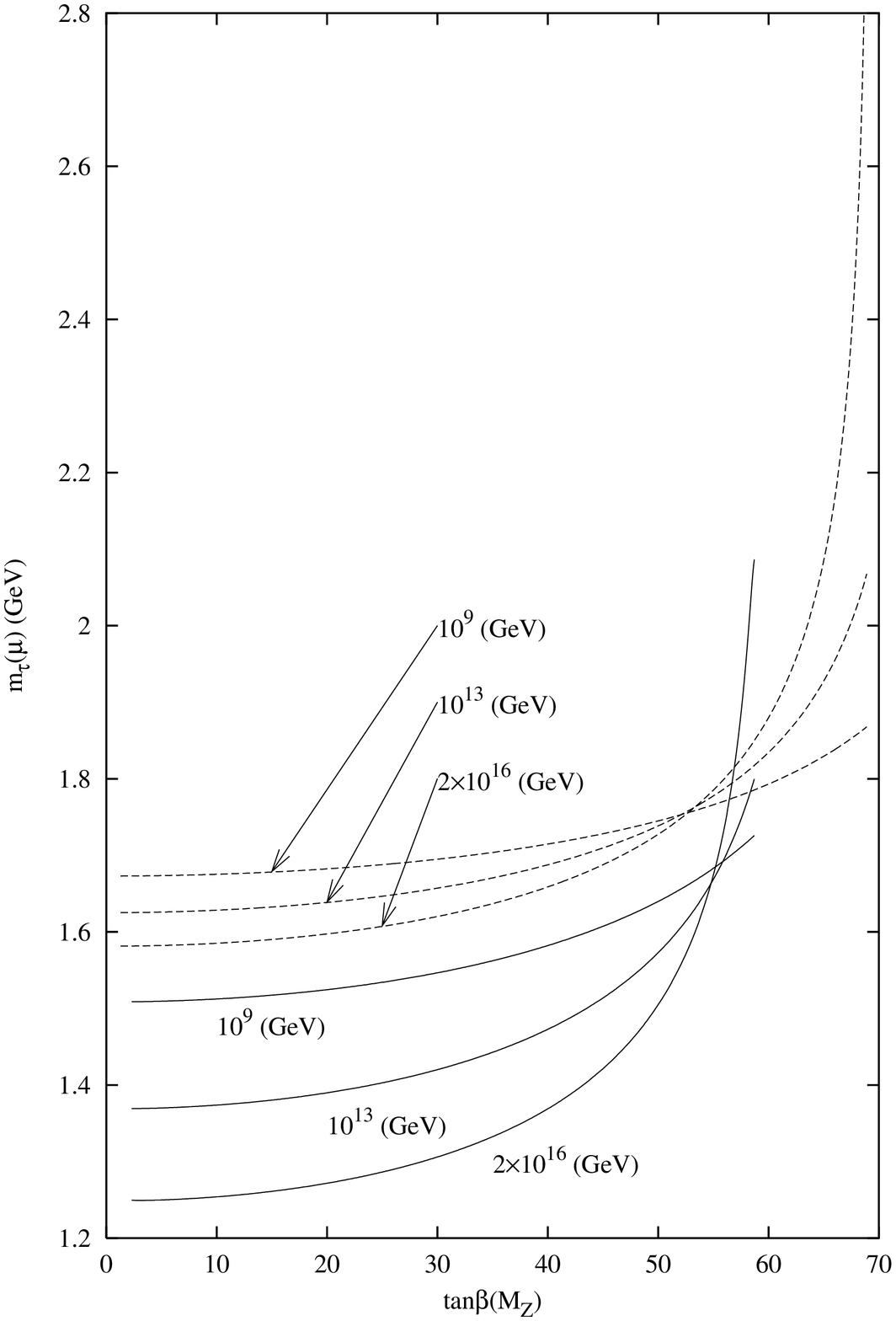}}
\caption{Predicton of $m_\tau(\mu)$ as at higher scales, $\mu=10^9$ GeV,
$10^{13}$ GeV and $2\times 10^{16}$ GeV as a function of $\tan\beta(M_Z)$
in MSSM with $M_S=M_Z$ (solid lined) and 2HDM (dashed lines).}
\label{fig13}
\end{figure}

Using the central values of $m_t(M_Z)$, $m_b(M_Z)$ and $m_\tau(M_Z)$ from
(\ref{eq37}), we have studied variation of $m_t(\mu)$, $m_b(\mu)$ and
$m_\tau(\mu)$ for different values of $\mu=10^9$ GeV, $10^{13}$ GeV and
$2\times 10^{16}$ GeV each as a function of various low-energy input values of
$\tan\beta(M_Z)$ in MSSM and 2HDM. These results are presented in
Figs.~\ref{fig11}-\ref{fig13} for the 2HDM (dashed lines) and for the MSSM
(solid lines) with $M_S=M_Z$. It is clear that the perturbatively allowed
range of $\tan\beta$ decreases with increasing $\mu$ both for MSSM and 2HDM.
\par
\begin{figure}
\epsfxsize=17cm
{\epsfbox{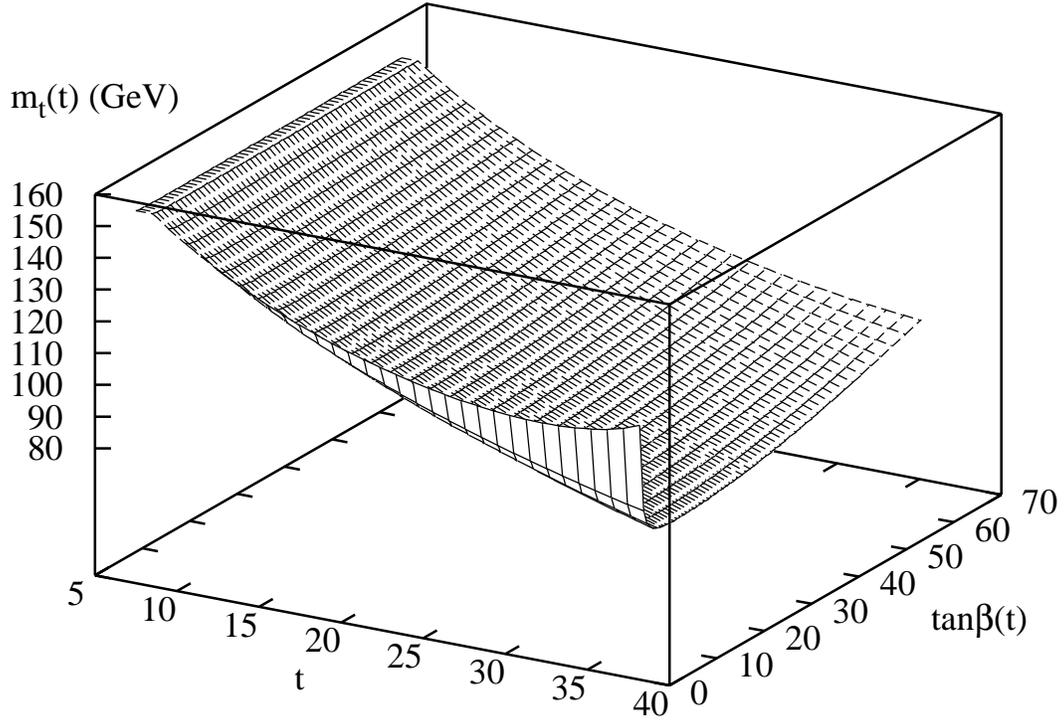}}
\caption{Prediction of top-quark mass $m_t(\mu)$ at higher scales $(\mu>M_Z)$
as a function of $\mu$ $(t=\ln\mu)$ and $\tan\beta(\mu)$ in MSSM with $M_S=1$
TeV. The values of $\tan\beta(\mu)$ at very $\mu$ has been obtained through
solutions of the corresponding RGE using $\tan\beta(M_Z)=2-58$ as inputs.}
\label{fig14}
\end{figure}
\begin{figure}
\epsfxsize=14cm
{\epsfbox{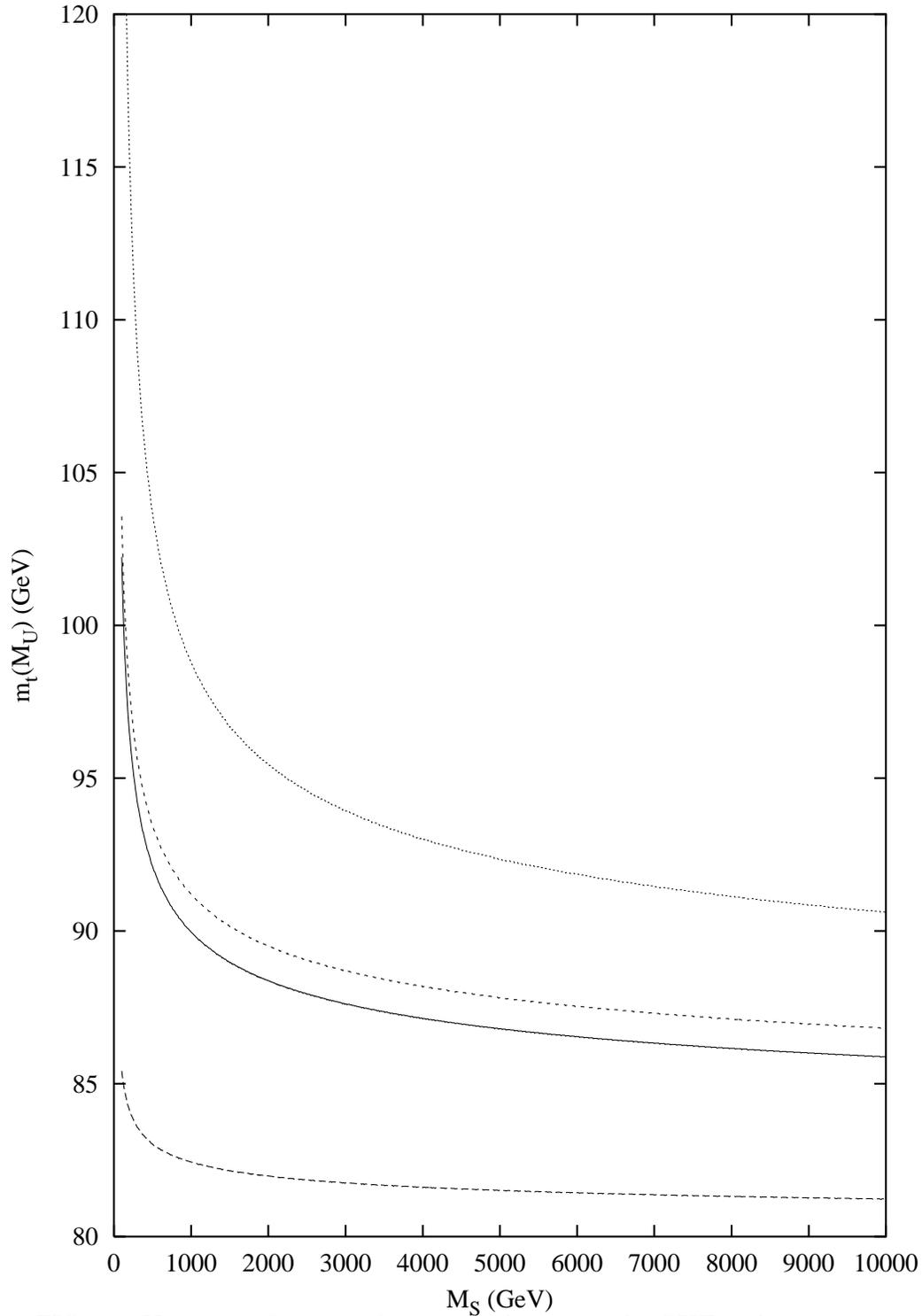}}
\caption{Variation of top-quark mass prediction at the GUT scale as a function
of SUSY scale $M_S=M_Z-10^{4}$ GeV and various values of $\tan\beta(M_S)$=3
(solid line), 10 (large-dashed line), 50 (small-dashed line), 55 (dotted
line).}
\label{fig15}
\end{figure}
\begin{figure}
\epsfxsize=14cm
{\epsfbox{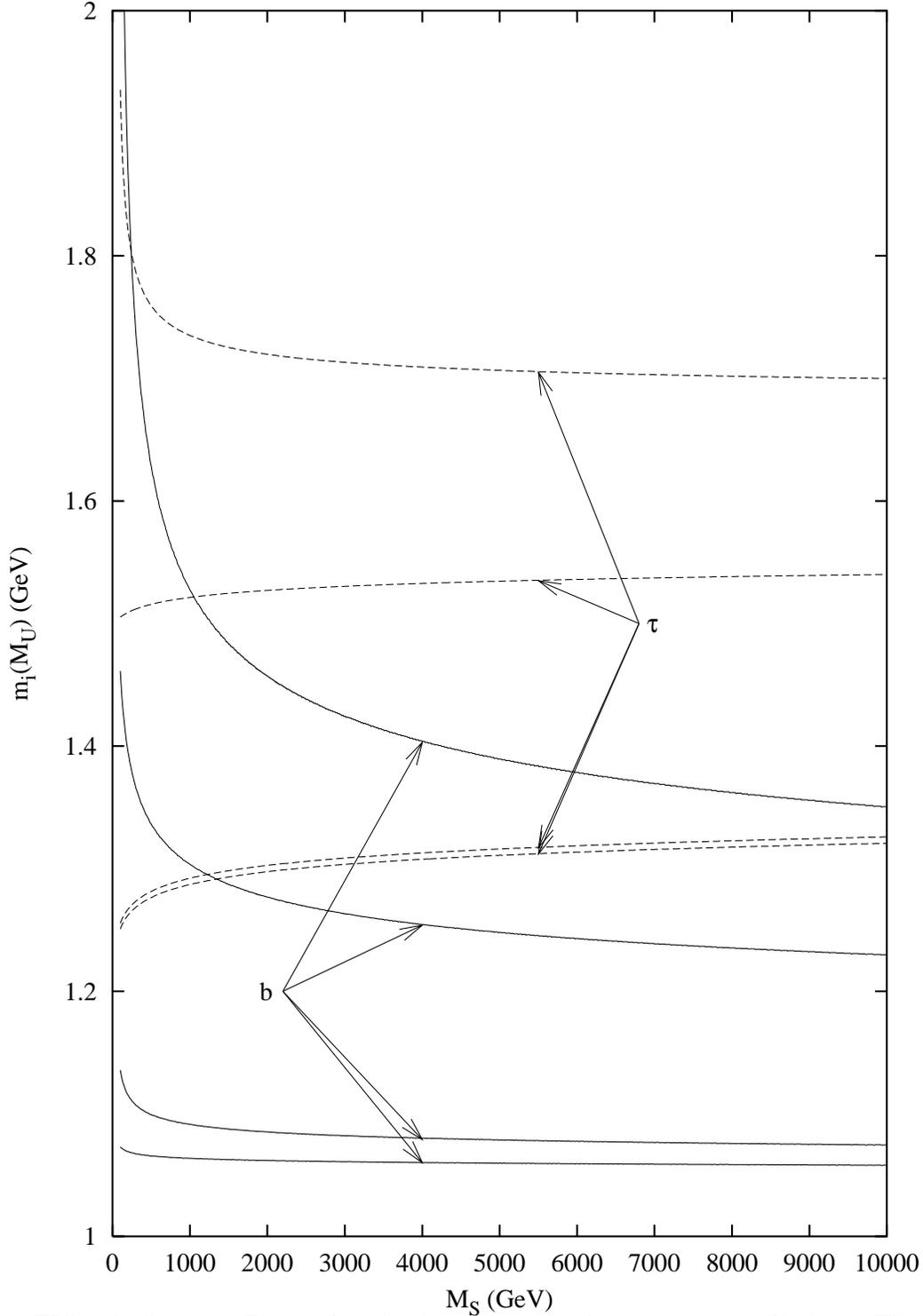}}
\caption{Same as Fig.~\ref{fig15} but for $b$-quark and $\tau$-lepton mass
predictions. The input values of $\tan\beta(M_S)$ for four different curves in
each case are $\tan\beta(M_S)=3,$ 10, 50, and 55 in the increasing order of
masses.}
\label{fig16}
\end{figure}

We have examined simultaneous variation of $m_t(\mu)$ as a function of $\mu$
and $\tan\beta(\mu)$ which is displayed in the three dimensional plot of
Fig.~\ref{fig14} for the input value of $m_t(M_Z)=181$ GeV and
$\tan\beta(M_S)=2-58$, where $M_S=1$ TeV. Using the top-quark mass at
$\mu=M_Z$, we have calculated $m_t(\mu)$ and $\tan\beta(\mu)$ at every $\mu$
between $M_S-M_U$ for input value of $\tan\beta(M_S)=2-59$. The results are
displayed in the three dimensional plot. The variations of the running mass
predictions at the GUT scale $(M_U=2\times 10^{16}$ GeV) as a function of the
SUSY scale $(M_S=M_Z-10^{4}$ GeV) are shown in Figs.~\ref{fig15}-\ref{fig16}
for the third generation fermions using various input values of
$\tan\beta(M_S)$. We find that the top-quark mass at the GUT scale at first
decreases sharply in the smaller and larger $\tan\beta$ regions as $M_S$
increases and then remains almost constant for $M_S=3\times 10^3-10^4$ GeV.
Similarly the predicted $b$ and $\tau$ masses decrease with increasing $M_S$
although fall off is slower in the case of $\tau$ in the large $\tan\beta$
region.
\par
Numerical values of predictions of the running masses are presented in
Table.~\ref{table1} for the SM at three different scales $\mu=10^9$ GeV,
$10^{13}$ GeV and $2\times 10^{16}$ GeV. The two-loop contributions to the RG
evolution of Yukawa couplings depends, although very weakly, upon the Higgs
quartic couplings $\lambda$ which is related to the Higgs mass $(M_H)$ and
VEV($v$), $\lambda=M_H^2/v^2$. We have used the two-loop RGEs for
$\lambda(\mu)$ for the SM \cite{16} and evaluated the running masses and VEVs
of Table.~\ref{table1} for the input value of the Higgs mass $M_H=250$ GeV.
Changing the Higgs mass in the allowed range of $M_H=220-260$ GeV \cite{32}
does not change the results of Table.~\ref{table1} significantly. The
uncertainties in the quantities are due to those in the running masses at
$\mu=M_Z$. The mass matrices for $M_b(\mu)$ and $M_u(\mu)$ are modified by the
factors $v(\mu)/v(M_Z)$ where $v(M_Z)\approx 174$ GeV and the CKM matrices at
higher scales remain the same as in \cite{32}. The computed values of masses
are found to be less when compared to those obtained with scale-independent
assumption \cite{32}. This is clearly understood as the running VEV in the SM
decreases with increasing $\mu$. For example in the SM at $\mu=2\times
10^{16}$ GeV our predictions are $(m_u, m_c, m_t)=(0.83$ MeV, 242.6 MeV,
75.4 GeV) as compared to \cite{32} $(m_u, m_c, m_t)=(0.94$ MeV, 272 MeV,
84 GeV) where the running effect on the VEV has been ignored. In Tables
\ref{table2}-\ref{table3} numerical values of masses, VEVs and $\tan\beta$
are given at the same three scales for the MSSM and 2HDM with
$\tan\beta(M_Z)=10$ and 55 in each of the two theories. As emphasized in this
paper our high scale estimations predict quite significantly different values
for the running masses especially in the up-quark sector. Although the CKM
matrices at high scales remain the same as in scale-independent assumptions
on the VEVs, the up-quark mass matrices are modified by the factor
$v_u(\mu)/v_u(M_S)$, but the down-quark and charged-lepton mass matrices are
modified by the factor $v_d(\mu)/v_d(M_S)$. In the MSSM with $M_S=M_Z$ and
$\tan\beta(M_Z)=10$ including running effect of the VEVs, the GUT scale
predictions are $(m_u, m_c, m_t)=(0.70$ MeV, 200 MeV, 73.5 GeV) as compared to
\cite{32} $(m_u, m_c, m_t)=(1.04$ MeV, 302 MeV, 129 GeV). But by increasing
the SUSY scale to $M_S=1$ TeV and in the large $\tan\beta$-region we find
substantial decrease in the predicted values of the top-quark mass at the GUT
scale leading to $(m_u, m_c, m_t)=(0.72$ MeV, 210 MeV, 95.1 GeV). This is
understood by noting that $\tan\beta\approx 55$ is closer to the perturbative
limit for which the top-quark Yukawa coupling is larger. Similarly from
Table.~\ref{table2} we note nearly a 20\% increase in the $m_b(M_U)$ with the
increase of $\tan\beta$ from $10-55$. Similar effects are also noted in 2HDM
as can be seen from Table.~\ref{table3} where $m_t$ decreases by nearly 7\%
as $\tan\beta$ increases from $10-55$. For larger effect the increase has to
be larger in $\tan\beta$ since the perturbative limit in this case is closer
to $\tan\beta\approx 69$ as compared to the MSSM case where the limiting value
is $\tan\beta\approx 59$.
\par
\section{Summary and Conclusion}
\label{sec6}
\par
In the frequently exploited renormalisation scheme in gauge theories, the
Yukawa couplings and VEVs in the SM, 2HDM and MSSM run separately
\cite{15,16,17,18,19,20,21,22,23,24,25,26,27,28,29}. The effect of scale
dependence of the VEVs has been ignored while deriving analytic formulas
\cite{28,30,31} and obtaining numerical predictions at higher scales for the
running masses of fermions \cite{32}, but appropriately taken into account
more recently \cite{13}. In this paper we have derived new analytic formulas
in the SM and 2HDM and generalised the formulas of \cite{13} for any
supersymmetry scale $(M_S>M_Z)$. The new formulas exhibit substantially
different functional dependence on gauge and Yukawa couplings in all the
three effective theories. In particular, the running masses of the first two
generations are found to be independent of Yukawa couplings of the third
generations in the small mixing limit. Numerical predictions at two-loop
level shows that all the running masses in the SM and only the up-quark
masses in the MSSM and 2HDM decrease at high scales when compared with the
predictions taking scale-independent VEVs. But in the case of MSSM and 2HDM
the down-quark and the charged-lepton masses increase over the corresponding
predictions obtained with scale-independent assumptions on the VEVs. Compared
to the MSSM the perturbatively allowed region of $\tan\beta$ is larger in
2HDM. In MSSM the allowed region shows a slow increase with the SUSY scale.
We have also made predictions of the running masses at the GUT scale as a
function of supersymmetry scale exhibiting new behaviours. We suggest that
these new analytic formulas and improved estimations on the running masses and
$\tan\beta$ at high scales be used as inputs to test models proposing unified
explanations of quark-lepton masses.
\par
\acknowledgments
\par
One of us (M.K.P.) thanks Professor E.A.~Paschos for hospitality at the
Institute of Physics, Dortmund University and Professor Y.~Achiman for
hospitality at the Institute of Physics, Wuppertal University where a part of
this work was carried out. The authors thank Professor L.~Lavoura for a useful
comment. This work is supported by the DAE project No.~98/37/91/BRNS-Cell/731
of the Govt.~of India.
\par

\par
\par
\par
\par
\par
\par
\par
\par
\par
\par
\par
\par
\par
\par
\par
\par
\par
\mediumtext
\begin{table}
\caption{Running mass and VEV predictions at higher scales in the nonSUSY
standard model for the input values of the Higgs mass $M_H=250$ GeV and other
parameters given in (\ref{eq37})-(\ref{eq39}).}
\begin{tabular}{|c|c|c|c|}\hline
&$\mu=10^9$ (GeV)&$\mu=10^{13}$ (GeV)&$\mu=2\times 10^{16}$ (GeV)\\\hline
$m_u$ (MeV)&$1.1537^{+0.2233}_{-0.2331}$&$0.9472^{+0.1849}_{-0.1923}$&
$0.8351^{+0.1636}_{-0.1700}$\\\hline
$m_c$ (MeV)&$335.2184^{+31.8261}_{-33.5603}$&$275.2419^{+26.5286}_{-27.8710}$&
$242.6476^{+23.5536}_{-24.7026}$\\\hline
$m_t$ (GeV)&$99.1359^{+10.7438}_{-9.8347}$&$83.9249^{+10.2622}_{-9.0281}$&
$75.4348^{+9.9647}_{-8.5401}$\\\hline
$m_d$ (MeV)&$2.3558^{+0.6513}_{-0.3538}$&$1.9529^{+0.5433}_{-0.2953}$&
$1.7372^{+0.4846}_{-0.2636}$\\\hline
$m_s$ (MeV)&$46.9155^{+6.5228}_{-6.9737}$&$38.8929^{+5.4652}_{-5.8228}$&
$34.5971^{+4.8857}_{-5.1971}$\\\hline
$m_b$ (GeV)&$1.3639^{+0.0328}_{-0.0398}$&$1.0971^{+0.0143}_{-0.0248}$&
$0.9574^{+0.0037}_{-0.0169}$\\\hline
$m_e$ (MeV)&$0.4665^{+0.0001}_{-0.0001}$&$0.4533^{+0.0001}_{-0.0001}$&
$0.4413^{+0.0001}_{-0.0001}$\\\hline
$m_\mu$ (MeV)&$98.4648^{+0.0049}_{-0.0050}$&$95.6834^{+0.0078}_{-0.0084}$&
$93.1431^{+0.0136}_{-0.0101}$\\\hline
$m_\tau$ (GeV)&$1.6738^{+0.0004}_{-0.0003}$&$1.6265^{+0.0005}_{-0.0004}$&
$1.5834^{+0.0001}_{-0.0005}$\\\hline
$v$ (GeV)&$157.5206^{-7.1815}_{+6.0558}$&$155.7062^{-10.6592}_{-8.5945}$&
$155.6196^{-13.6336}_{+10.4664}$\\\hline
\end{tabular}
\label{table1}
\end{table}
\par
\begin{table}
\caption{Predictions of running masses, VEVs and $\tan\beta$ at higher scales
$\mu=10^9$ GeV, $10^{13}$ GeV and $2\times 10^{16}$ GeV in MSSM with SUSY
scale $M_S=1$ TeV, using two-loop RG equations.}
\begin{tabular}{|c|c|c|c|}\hline
$\tan\beta(M_S)=10$&$\mu=10^9$ (GeV)&$\mu=10^{13}$ (GeV)&$\mu=2\times
10^{16}$ (GeV)\\\hline
$m_u$ (MeV)&$1.1618^{+0.2226}_{-0.2345}$&$0.8882^{+0.1694}_{-0.1794}$&
$0.7238^{+0.1365}_{-0.1467}$\\\hline
$m_c$ (MeV)&$339.4064^{+31.2929}_{-33.4804}$&$258.0945^{+23.8287}_{-25.8339}$&
$210.3273^{+19.0036}_{-21.2264}$\\\hline
$m_t$ (GeV)&$112.3144^{+17.0392}_{-13.7215}$&$94.3698^{+22.5577}_{-14.4831}$&
$82.4333^{+30.2676}_{-14.7686}$\\\hline
$m_d$ (MeV)&$2.3842^{+0.6582}_{-0.3574}$&$1.8290^{+0.5111}_{-0.2779}$&
$1.5036^{+0.4235}_{-0.2304}$\\\hline
$m_s$ (MeV)&$47.4812^{+6.5845}_{-7.0454}$&$36.4261^{+5.1588}_{-5.4807}$&
$29.9454^{+4.3001}_{-4.5444}$\\\hline
$m_b$ (GeV)&$1.5920^{+0.1038}_{-0.0915}$&$1.2637^{+0.1189}_{-0.0893}$&
$1.0636^{+0.1414}_{-0.0865}$\\\hline
$m_e$ (MeV)&$0.4290^{+0.0001}_{-0.0001}$&$0.3911^{+0.0002}_{-0.0002}$&
$0.3585^{+0.0003}_{-0.0003}$\\\hline
$m_\mu$ (MeV)&$90.5439^{+0.0169}_{-0.0173}$&$82.5539^{+0.0346}_{-0.0330}$&
$75.6715^{+0.0578}_{-0.0501}$\\\hline
$m_\tau$ (GeV)&$1.5429^{+0.0006}_{-0.0006}$&$1.4085^{+0.0009}_{-0.0008}$&
$1.2922^{+0.0013}_{-0.0012}$\\\hline
$\tan\beta$&$8.2314^{-0.5046}_{+0.3807}$&$7.4350^{-0.9752}_{+0.6302}$&
$6.9280^{-1.5156}_{+0.8234}$\\\hline
$v_u$ (GeV)&$141.7765^{-9.7365}_{+7.6253}$&$130.5455^{-18.0431}_{+12.1155}$&
$123.8177^{-27.8954}_{+15.7651}$\\\hline
$v_d$ (GeV)&$17.2237^{-0.1352}_{+0.1241}$&$17.5581^{-0.1426}_{+0.1302}$&
$17.8718^{-0.1492}_{+0.1354}$\\\hline\hline
$\tan\beta(M_S)=55$&$\mu=10^9$ (GeV)&$\mu=10^{13}$ (GeV)&
$\mu=2\times 10^{16}$ (GeV)\\\hline
$m_u$ (MeV)&$1.1687^{+0.2225}_{-0.2346}$&$0.8889^{+0.1675}_{-0.1795}$&
$0.7244^{+0.1219}_{-0.1466}$\\\hline
$m_c$ (MeV)&$339.5917^{+31.2621}_{-33.5026}$&$258.2929^{+23.3295}_{-25.8144}$&
$210.5049^{+15.1077}_{-21.1538}$\\\hline
$m_t$ (GeV)&$118.6588^{+19.9035}_{-15.4790}$&$104.2363^{+32.7015}_{-18.2028}$&
$95.1486^{+69.2836}_{-20.659}$\\\hline
$m_d$ (MeV)&$2.3774^{+0.6542}_{-0.3553}$&$1.8219^{+0.5054}_{-0.2755}$&
$1.4967^{+0.4157}_{0.2278}$\\\hline
$m_s$ (MeV)&$47.3523^{+6.5303}_{-7.0069}$&$36.2891^{+5.0777}_{-5.4340}$&
$29.8135^{+4.1795}_{-4.4967}$\\\hline
$m_b$ (GeV)&$1.8297^{+0.1667}_{-0.1376}$&$1.5768^{+0.2640}_{-0.1685}$&
$1.4167^{+0.4803}_{-0.1944}$\\\hline
$m_e$ (MeV)&$0.4276^{-0.0003}_{+0.0001}$&$0.3893^{-0.0005}_{+0.0002}$&
$0.3565^{-0.001}_{+0.0002}$\\\hline
$m_\mu$ (MeV)&$90.2779^{-0.0508}_{+0.0318}$&$82.2064^{-0.1024}_{+0.0468}$&
$75.2938^{-0.1912}_{+0.0515}$\\\hline
$m_\tau$ (GeV)&$1.6867^{+0.0056}_{-0.005}$&$1.6574^{+0.0188}_{-0.0148}$&
$1.6292^{+0.0443}_{-0.0294}$\\\hline
$\tan\beta$&$53.6122^{-2.3644}_{+1.5356}$&$52.7633^{-6.3597}_{+2.9538}$&
$52.0738^{-16.5475}_{+4.3757}$\\\hline
$v_u$ (GeV)&$141.2095^{-10.6285}_{+8.1355}$&$127.4742^{-22.6973}_{+13.8538}$&
$117.7947^{-46.7214}_{+19.2752}$\\\hline
$v_d$ (GeV)&$2.6339^{-0.0859}_{+0.0741}$&$2.4159^{-0.158}_{+0.1206}$&
$2.2620^{-0.2615}_{+0.1661}$\\\hline
\end{tabular}
\label{table2}
\end{table}
\par
\begin{table}
\caption{Predictions of running masses, VEVs and $\tan\beta$ in 2HDM at higher
scales using one-loop RG equations.}
\begin{tabular}{|c|c|c|c|}\hline
$\tan\beta(M_S)=10$&$\mu=10^9$ (GeV)&$\mu=10^{13}$ (GeV)&$\mu=2\times 10^{16}$
(GeV)\\\hline
$m_u$ (MeV)&$1.2021^{+0.2309}_{-0.2417}$&$0.9908^{+0.1919}_{-0.2002}$&
$0.8749^{+0.1701}_{-0.1772}$\\\hline
$m_c$ (MeV)&$349.2805^{+32.6824}_{-34.5798}$&$287.8975^{+27.3606}_{-28.8305}$&
$254.2131^{+24.3398}_{-25.5998}$\\\hline
$m_t$ (GeV)&$103.5011^{+11.3400}_{-10.2307}$&$88.2332^{+11.1753}_{-9.5397}$&
$79.6373^{+11.1974}_{-9.127}$\\\hline
$m_d$ (MeV)&$2.4547^{+0.6748}_{-0.366}$&$2.0430^{+0.5650}_{-0.3069}$&
$1.8204^{+0.505}_{-0.2743}$\\\hline
$m_s$ (MeV)&$48.8852^{+6.7278}_{-7.2144}$&$40.6860^{+5.6602}_{-6.0484}$&
$36.2544^{+5.0700}_{-5.4083}$\\\hline
$m_b$ (GeV)&$1.6281^{+0.0910}_{-0.0854}$&$1.3709^{+0.0854}_{-0.0775}$&
$1.2309^{+0.0826}_{-0.0730}$\\\hline
$m_e$ (MeV)&$0.4662^{+0.0001}_{-0.0001}$&$0.4529^{+0.0001}_{-0.0001}$&
$0.4407^{+0.0001}_{-0.0001}$\\\hline
$m_\mu$ (MeV)&$98.4132^{+0.0050}_{-0.0051}$&$95.5970^{+0.0086}_{-0.0086}$&
$93.0197^{+0.0122}_{-0.0122}$\\\hline
$m_\tau$ (GeV)&$1.6752^{+0.0004}_{-0.0004}$&$1.6283^{+0.0004}_{-0.0004}$&
$1.5851^{+0.0005}_{-0.0005}$\\\hline
$\tan\beta$&$8.1956^{-0.3894}_{+0.3255}$&$7.6757^{-0.5649}_{+0.4496}$&
$7.3543^{-0.6975}_{+0.5348}$\\\hline
$v_u$ (GeV)&$155.6481^{-7.4729}_{+6.2622}$&$152.8315^{-11.3442}_{+9.0595}$&
$151.9551^{-14.5219}_{+11.1741}$\\\hline
$v_d$ (GeV)&$18.9914^{-0.0098}_{+0.0095}$&$19.9110^{-0.0137}_{+0.0131}$&
$20.6620^{-0.0167}_{+0.0157}$\\\hline\hline
$\tan\beta(M_S)=55$&$\mu=10^9$ (GeV)&$\mu=10^{13}$ (GeV)&
$\mu=2\times 10^{16}$ (GeV)\\\hline
$m_u$ (MeV)&$1.2021^{+0.2309}_{-0.2417}$&$0.9908^{+0.1919}_{-0.2002}$&
$0.8749^{+0.1701}_{-0.1772}$\\\hline
$m_c$ (MeV)&$349.2889^{+32.6905}_{-34.5782}$&$287.9066^{+27.3646}_{-28.8338}$&
$254.2223^{+24.3441}_{-25.6031}$\\\hline
$m_t$ (GeV)&$106.6700^{+12.2719}_{-10.9231}$&$92.1000^{+12.6648}_{-10.5174}$&
$83.9317^{+13.2279}_{-10.3226}$\\\hline
$m_d$ (MeV)&$2.4547^{+0.6748}_{-0.3660}$&$2.0430^{+0.5650}_{-0.3069}$&
$1.8204^{+0.5050}_{-0.2743}$\\\hline
$m_s$ (MeV)&$48.8888^{+6.7295}_{-7.2159}$&$40.6898^{+5.6622}_{-6.0498}$&
$36.2584^{+5.0720}_{-5.4099}$\\\hline
$m_b$ (GeV)&$1.7719^{+0.1203}_{-0.1092}$&$1.5392^{+0.1272}_{-0.1092}$&
$1.4128^{+0.1353}_{-0.1162}$\\\hline
$m_e$ (MeV)&$0.4662^{+0.0001}_{-0.0001}$&$0.4529^{+0.0001}_{-0.0001}$&
$0.4407^{+0.0001}_{0.0001}$\\\hline
$m_\mu$ (MeV)&$98.4302^{+0.0054}_{-0.0055}$&$95.6235^{+0.0097}_{-0.0094}$&
$93.0536^{+0.0146}_{-0.0136}$\\\hline
$m_\tau$ (GeV)&$1.7659^{+0.0028}_{-0.0025}$&$1.7775^{+0.0073}_{-0.0060}$&
$1.7851^{+0.0136}_{-0.0107}$\\\hline
$\tan\beta$&$54.9963^{-1.5534}_{+1.1589}$&$55.7094^{-2.4787}_{+1.6588}$&
$56.5831^{-3.2730}_{+1.9895}$\\\hline
$v_u$ (GeV)&$155.7178^{-7.8644}_{+6.5265}$&$152.0846^{-12.3141}_{+9.6439}$&
$150.4478^{-16.1914}_{+12.0879}$\\\hline
$v_d$ (GeV)&$2.8314^{-0.0649}_{+0.0578}$&$2.7299^{-0.1042}_{+0.0892}$&
$2.6588^{-0.1404}_{+0.1161}$\\\hline
\end{tabular}
\label{table3}
\end{table}
\narrowtext
\end{document}